\documentclass[aoas,preprint]{imsart}

\RequirePackage[OT1]{fontenc}
\RequirePackage{amsthm,amsmath}
\RequirePackage{natbib}
\usepackage[utf8]{inputenc}
\usepackage{amsmath,natbib}
\usepackage{amsthm}
\usepackage{amsfonts}
\usepackage{mathtools}
\usepackage{amssymb}
\usepackage{graphicx}
\usepackage{float}
\usepackage{pgfplotstable}
\usepackage{booktabs}
\usepackage{colortbl}
\usepackage{multirow}
\usepackage{array}
\usepackage{longtable}
\usepackage{tabu}
\usepackage{rotating}
\usepackage{lscape}
\usepackage{afterpage}
\usepackage{multirow}
\usepackage{subfigure}
\usepackage{listings}
\usepackage{color}

\newcommand\norm[1]{\left\lVert#1\right\rVert}

\startlocaldefs
\numberwithin{equation}{section}
\theoremstyle{plain}

\endlocaldefs

\begin{document}

\begin{frontmatter}
\title{Sparse Mean-Variance Portfolios: A Penalized Utility Approach}
\runtitle{Sparse Mean-Variance Portfolios}

\begin{aug}
\author{\fnms{David} \snm{Puelz}\ead[label=e1]{david.puelz@utexas.edu}},
\author{\fnms{P. Richard} \snm{Hahn}\ead[label=e2]{richard.hahn@chicagobooth.edu}}
\and
\author{\fnms{Carlos M.} \snm{Carvalho}
\ead[label=e3]{carlos.carvalho@mccombs.utexas.edu}}
\runauthor{Puelz, Hahn and Carvalho}

\affiliation{The University of Texas and The University of Chicago}

\address{David Puelz\\
\printead{e1}\\}

\address{P. Richard Hahn\\
\printead{e2}\\}

\address{Carlos M. Carvalho\\
\printead{e3}\\}

\end{aug}

\begin{abstract}
This paper considers mean-variance optimization under uncertainty, specifically when one desires a sparsified set of optimal portfolio weights.  From the standpoint of a Bayesian investor, our approach produces a small portfolio from many potential assets while acknowledging uncertainty in asset returns and parameter estimates.  We demonstrate the procedure using static and dynamic models for asset returns.

\end{abstract}

%

\end{frontmatter}

\section{Introduction}

Practical investing requires balancing portfolio optimality and simplicity. In other words, investors desire well-performing portfolios that are \textit{easy} to manage, and this preference is driven by many factors. Managing large asset positions and transacting frequently is expensive and time consuming, and these complications arise from both trading costs and number of assets available for investment.  For the individual investor, these challenges are strikingly amplified. Their choice set for investment opportunities is massive and includes exchange traded funds (ETFs), mutual funds, and thousands of individual stocks.  Further, transaction costs can become prohibitively large even for a modest number of orders \citep{barber2000trading,barber2009just}. This raises the question: \textit{How does one invest optimally while keeping the simplicity of a portfolio in mind?} We tackle this question from a new statistical and decision-theoretic perspective.  Portfolio optimization requires inputs that are parameters from a statistical model (typically the mean and covariance of asset returns) which are inherently uncertain, and a Bayesian estimation of these parameters yields posterior distributions quantifying this uncertainty.  Because the optimal portfolio depends on these parameters \citep{markowitz1952portfolio}, the optimality criterion (typically characterized by a ratio of portfolio return and risk and known as the Sharpe ratio) is also uncertain (see \cite{PandV} and \cite{jacquier2012asset} for important Bayesian treatments of uncertainty in finance).  In this paper, we propose a way to study how much ``optimality" we lose ex ante for varying levels of portfolio simplicity.  We compare simple portfolios with their more complex counterparts while acknowledging statistical uncertainty.  One key result of our approach is that very little ex ante is lost (in terms of Sharpe ratio) by an optimal portfolio built on a small subset of assets when many assets are available for investment. We also find (through out-of-sample tests) that the ex post performance of these ``small portfolios" exceed that of their ``full portfolio" counterparts.  In other words, if the deterioration in the predicitive Sharpe ratio for increasingly simpler portfolios is negligible up to statistical uncertainty \textit{and} they perform well out-of-sample, why not hold a simpler and easier to manage portfolio?

Methodologically, our approach adapts the \textit{decoupling shrinkage and selection (DSS)} procedure from \cite{HahnCarvalho} to the portfolio selection problem (see also \cite{puelz2016variable}).  Their paper defines a new framework for model selection where statistical inference, regardless of prior and likelihood specifications, is separated from the model selection process itself. In general, the \textit{DSS} paradigm involves a model fitting step (inference) and a selection step (optimization). For the selection step, one specifies a loss function on which a particular model is graded, often by predictive loss.  The two steps are married by integration over uncertainty, and the final loss function is derived by integrating over the predictive and posterior distributions of the future observables and model parameters from the inference, respectively.  Then, this integrated loss function can be optimized. \cite{HahnCarvalho} develop the \textit{DSS} procedure for linear and gaussian graphical models and use variation explained and excess error heuristics for model selection. Extending this work, we apply the \textit{DSS} approach to the problem of portfolio selection and optimization.  The portfolio optimization loss function is defined as the portfolio return mean minus the portfolio return variance, and the portfolio Sharpe ratio is chosen as the model selection heuristic.   
 
In this paper, we focus on the problem of investing in exchange traded funds (ETFs). ETFs have gained popularity over the past two decades as a low-cost way for individual investors to hold large baskets of stocks.  These funds are mandated to closely follow (track) the returns of stock market indices.  For example, the ETF SPY tracks the S\&P 500 index which is comprised of 500 large market-capitalization companies.  An individual investor has increasing numbers of ETFs to choose from, including funds that invest in currencies, foreign equities, bonds, real estate, and commodities.  Given this large choice set and modest means by which individuals can invest, we utilize our approach to optimally select a small subset of ETFs.  Several financial startups are now providing ``ETF selection" services, including Betterment and Wealthfront.  After answering a series of questions to determine an investor's risk tolerance, these companies provide small portfolios of ETFs for investment.  ETF selection and investing is clearly a relevant problem for individual investors today.



\subsection{Previous research}
There are two streams of research in portfolio optimization that are important to this work.  The first focuses on the portfolio selection problem, particularly in stock investing and index tracking.  \cite{polson1999bayesian} consider the S\&P 500 index and develop a Bayesian approach for large-scale stock selection and portfolio optimization from the index's constituents.  Other insightful Bayesian approaches to optimal portfolio choice in different contexts include \cite{johannes2014sequential}, \cite{gron2012optimal}, \cite{jacquier2010simulation}, \cite{puelz2015optimal} and \cite{pettenuzzo2015optimal}.  

A second stream of research is dedicated to regularization techniques in mean-variance optimization. Mean-variance portfolio optimization is a cornerstone of finance where optimal weights are found by maximizing portfolio return for a fixed level of risk \citep{markowitz1952portfolio}.  Since this procedure relies on estimates of the first two moments of return, the optimal weights depend heavily on the modeling choice for these moments.  It is well documented that sample estimates for the mean and variance produce portfolios with extreme weights that perform poorly out-of-sample \citep{jobson1980estimation}, \citep{best1991sensitivity},  \citep{broadie1993computing}, \citep{britten1999sampling}, \citep{demiguel2009optimal}, \citep{frankfurter1971portfolio}, \citep{dickinson1974reliability}, and \citep{frost1988better}.  Estimation errors are the culprit, and it is widely acknowledged that errors in the expected return are much larger than errors in the covariance matrix, see for example \cite{merton1980estimating}.  As a result, researchers have focused on ways to stabilize, through regularization, the entire optimization process. For a detailed literature review of finance literature related to regularized portfolio optimization, we refer the reader to the appendix.

Under a new decision-theoretic lens, this paper frames weight regularization in a portfolio selection methodology by viewing it as a necessary part of the investor's utility.  This is markedly different from previous literature where weights are regularized for statistical benefits -- thus mixing investor preferences with the modeling of the optimization inputs.  Our loss function regularizes the weights to mirror the investor's joint preference for portfolio \textit{simplicity} and high risk-adjusted return, and we are able to tailor the utility to many other desires of the investor (see section 4).  The paradigm of \textit{decoupling shrinkage and selection} used in this paper is intended to clearly separate statistical modeling from the investor's utility optimization problem.  In this way, we disentangle the motivation for using a penalized utility from everything statistical -- a contribution that does not currently exist in the literature.  Additionally, we show financial benefits of this new method in a practical out-of-sample exercise in section \ref{DLMsection}. 



\subsection{Outline} The paper is structured as follows.  In section 2, we derive a loss function for a mean-variance investor to be used for portfolio choice.  In section 3, we demonstrate how this loss function may be used to construct portfolios of ETFs. 

The first part of section 3 is an illustration of the procedure -- we consider two static models for asset returns (a latent factor model and normal-inverse-Wishart model) to emphasize how the methodology is agnostic to choice of prior and likelihood. Since the \textit{DSS} framework permits the separation of the \textit{learning stage} from the \textit{decision stage}, we show how two reasonable choices for learning the moments of asset returns propagate into the overall portfolio decision.

The second part of section 3 considers a practical application of the methodology anchored in the ideas from the illustrative section.  We construct a dynamic model for asset returns and show how an investor may implement an out-of-sample portfolio strategy which performs well compared to several alternatives.

\section{Loss function derivation} \label{LossFunctionDerivation}

We consider a portfolio optimization involving the first two moments of portfolio return.  Assume we desire to invest in a set of $N$ assets with stochastic returns given by the vector $\tilde{R}$.  We consider long only portfolios since layman investors are often unable hold short positions.  We consider the following optimization:

\begin{equation} \label{optprob}
	\begin{split}
	\max_{w \geq 0} \hspace{3mm} \mathbb{E}_{p(\tilde{R})}\left[\sum_{k=1}^{N} w_{k}\tilde{R}_{k} - \frac{1}{2}\sum_{k=1}^{N}\sum_{j=1}^{N} w_{k}w_{j}\tilde{R}_{k}\tilde{R}_{j} + \lambda \norm{ w }_{1} \right],
	\end{split}
\end{equation}where $w$ is vector giving the amount invested in each asset.  The first two terms in objective \ref{optprob} is the second order Taylor approximation to the exponential growth rate of wealth expanded about $w_{0} = \vec{0}$.  Assuming returns and weights are small enough, this is a reasonable approximation.  This can also be viewed as a version of the Kelly portfolio criterion where the mean and covariance of the asset returns appear (see \cite{kelly1956new} for the original paper).  Also, this approximation gives the maximum Sharpe ratio portfolio which is a unique among all possible optimal portfolios on the efficient frontier (i.e. the set of optimal portfolios for varying levels of risk).  The third term in objective \ref{optprob} is a penalty on the portfolio weights.  This reflects the investor's desire for portfolio simplicity and is not used for regularization of poor inference as previous studies have done.  Intuitively, this penalty ``sparsifies" the optimal weight vector by allocating a reasonable amount of wealth to a small subset of assets.    

Assume that $\tilde{R}$ is generated from a distribution with mean $\mu$ and variance $\Sigma$:

\begin{equation}
	\begin{split}
	\tilde{R} \sim \Pi(\mu,\Sigma).
	\end{split}
\end{equation}Given this assumption, we pass the expectations through the objective in problem \ref{optprob} to obtain:

\begin{equation} \label{optprob2}
	\begin{split}
	\max_{w \geq 0} \hspace{3mm} w^{T}\mu - \frac{1}{2}w^{T}\Sigma w + \lambda \norm{ w }_{1}.
	\end{split}
\end{equation}Note that the expectations avoid the penalty term on the weights.

In the context of \cite{HahnCarvalho}, we define the integrated loss function from the objective in problem \ref{optprob2} as:

\begin{equation}
	\begin{split}
		\mathcal{L}(w,\mu,\Sigma) = \frac{1}{2}w^{T}\Sigma w - w^{T}\mu + \lambda \norm{ w }_{1}.
	\end{split}
\end{equation}Suppose we have a posterior distribution for the model parameters $\mu$ and $\Sigma$.  As a final step, we integrate over this distribution to obtain:

\begin{equation}
	\begin{split}
	 \mathcal{L}(w) &= \mathbb{E}_{p(\mu,\Sigma)}\left[ \mathcal{L}(w,\mu,\Sigma) \right]
	 \\
	 &=\frac{1}{2}w^{T}\overline{\Sigma} w - w^{T}\overline{\mu} + \lambda \norm{ w }_{1},
	\end{split}
\end{equation} since $ \mathbb{E}[w^{T} \Sigma w] =  \text{tr}\left(\mathbb{E}[w^{T} \Sigma w]\right) =  \mathbb{E}[\text{tr}\left(w w^{T} \Sigma \right) ] = \text{tr}(w w^{T} \overline{\Sigma} ) = w^{T}\overline{\Sigma} w$.  Completing the square in $w$, dropping terms that do not involve the portfolio weights, and defining $\overline{\Sigma} = LL^{T}$, we rewrite the first two terms of the loss function as an $l_{2}$-norm:

 \begin{equation}
	\begin{split}
	 \mathcal{L}(w) &=\frac{1}{2}w^{T}\overline{\Sigma} w - w^{T}\overline{\mu} + \lambda \norm{ w }_{1}
	 \\
	 &\propto \frac{1}{2}(w-\overline{\Sigma}^{-1}\overline{\mu})^{T}\overline{\Sigma}(w-\overline{\Sigma}^{-1}\overline{\mu}) + \lambda \norm{ w }_{1}
	 \\
	 &= \frac{1}{2}(w-L^{-T}L^{-1}\overline{\mu})^{T}LL^{T}(w-L^{-T}L^{-1}\overline{\mu}) + \lambda \norm{ w }_{1}
	 \\
	 &= \frac{1}{2} \left[ L^{T}(w-L^{-T}L^{-1}\overline{\mu}) \right]^{T} \left[ L^{T}(w-L^{-T}L^{-1}\overline{\mu}) \right] + \lambda \norm{ w }_{1}
	 \\
	 &= \frac{1}{2} \norm{ L^{T}w-L^{-1}\overline{\mu} }_{2}^{2} + \lambda \norm{ w }_{1}.
	\end{split}
\end{equation} The optimization of interest is written as:

\begin{equation}
	\begin{split} \label{optprobfinal}
		 \min_{w \geq 0} \hspace{3mm} \frac{1}{2} \norm{ L^{T}w-L^{-1}\overline{\mu} }_{2}^{2} + \lambda \norm{ w }_{1}.
	\end{split}
\end{equation}

After integrating the loss function over the predictive distribution, notice how portfolio optimality and simplicity appear in optimization \ref{optprobfinal}.  The $l_{2}$ norm includes the term, $L^{-1}\overline{\mu}$ and the weights are scaled by $L^{T}$, the right Cholesky factor of the posterior mean of the covariance.  Optimizing the first term is similar to minimizing the distance between the chosen and maximum Sharpe ratio portfolios\footnote{In the unconstrained optimization, the maximum Sharpe ratio portfolio weights are given by $w^{*} \propto \overline{\Sigma}^{-1}\overline{\mu}$.}.  The simplicity component appears in the $l_1$ norm on the weights.  As previously mentioned, this penalty zeros out asset weights for larger value of the tuning parameter, $\lambda$.  Optimization \ref{optprobfinal} is now in the form of standard sparse regression loss functions, \citep{Tib}, with covariates, $L^{T}$, data, $L^{-1}\overline{\mu}$, and regression coefficients, $w$.  We may optimize \ref{optprobfinal} conveniently using existing software, such as the \texttt{lars} package of \cite{Efron}.

Choice of the $\lambda$ parameter is an important practical issue.  To this end, we employ the Bayesian approach adapted from \cite{HahnCarvalho} by examining plots showing the predictive deterioration attributable to $\lambda$-induced sparsification.  These plots convey the posterior uncertainty in any posterior metric chosen.  For this paper, we consider the Sharpe ratio, a portfolio performance metric of return divided by risk and first studied in \cite{sharpe1966mutual}.  The intuition of this heuristic can be articulated as follows (for instance): choose $\lambda$ such that, with posterior probability 95\%, the predictive deterioration in the sparsified portfolio's Sharpe ratio is no more than 10\% worse than the unsparsified optimal portfolio's Sharpe ratio.  The Sharpe ratio is a natural metric to consider since its inputs, the portfolio mean and variance, are embedded in utility function \ref{optprob} which leads to our final loss function.  

Selection of $\lambda$ based on predictive Sharpe ratio distribution plots is a useful financial heuristic but is not the only metric that may be chosen.  Any other functions of the moments of returns (and their posteriors) may be used.  The important idea borrowed from \cite{HahnCarvalho} is the novel use of statistical uncertainty for selection.

\section{Empirical Findings}
We now apply our derived loss function to ETF portfolio selection and optimization.  The empirical results are divided into two sections:

\begin{enumerate}
	\item Illustration of the procedure.
	\item Practical application of the procedure.
\end{enumerate}The first section illustrates how the portfolio selection procedure works by considering two models for the moments $\mu$ and $\Sigma$: (\textit{i}) Latent factor model and (\textit{ii}) Normal-inverse-Wishart model.  The second section builds on the first by developing a dynamic portfolio selection procedure relying on a dynamic regression model for $\mu_{t}$ and $\Sigma_{t}$. This is intended to explore the realistic scenario of an individual investor who desires to update their portfolio over time.

Specifically, the two models in the first section may be used for constructing \textit{static} or ``buy-and-hold" portfolios and illustrates the \textit{decoupling shrinkage and selection} procedure applied to the portfolio problem.  They are both shown to demonstrate how the portfolio decision may change under different specifications of prior and likelihood.  In contrast, the dynamic model provides estimations for time-varying moments and permits constructing a sequence of portfolios through time as more asset return data is observed.

We use data on the 25 most highly traded (i.e., most liquid) equity funds from ETFdb.com. This is monthly data from the Center for Research in Security Prices (CRSP) database from February 1992 through February 2015 \citep{CRSP}.  
In each example, we minimize the derived loss function,

\begin{equation}
	\begin{split} \label{optprobfinal2}
		 \mathcal{L}(w) = \frac{1}{2} \norm{ L^{T}w-L^{-1}\overline{\mu} }_{2}^{2} + \lambda \norm{ w }_{1},
	\end{split}
\end{equation} with a constraint on short-selling the portfolio assets and along the entire solution path by varying $\lambda$.  For the time-varying model, we substitute in the time dependent moment estimations and optimize \ref{optprobfinal2} at each time step. We consider an investor who is constrained to long positions in their portfolio (although short positions may be allowed by easily modifying the investor's  utility optimization).  We focus on a long only empirical analysis since ETF investors are typically leverage constrained.


\subsection{Illustration of the procedure} In this section, we show empirical results for a latent factor model and normal-inverse-Wishart model.  We present these two models in sequence to emphasize the flexibility of the procedure to choice of prior and likelihood and to compare how these choices affect the final portfolio decision.  Neither model is necessarily the \textit{``correct"} choice. Instead, both are shown to highlight the modularity of the methodology.

\subsubsection{Latent factor model}

\paragraph{Model overview}

We first consider a latent factor model for asset returns of the form:

\begin{equation}
	\begin{split}
		\tilde{R}_{t} &= \mu + \textbf{B}\textbf{f}_{t} + \textbf{v}_{t}
		\\
		&\textbf{v}_{t} \sim \text{N}(0,\mathbf{\Psi})
		\\
		&\textbf{f}_{t} \sim \text{N}(0,\textbf{I}_{k})
		\\
		&\mu \sim \text{N}(0,\Phi),
	\end{split}
\end{equation}where $\Psi$ is assumed diagonal and the set of $k$ latent factors $\textbf{f}_{t}$ are independent.  The covariance of the asset returns is constrained by the factor decomposition and takes the form:

\begin{equation}
	\begin{split}
		\Sigma = \textbf{B}\textbf{B}^{T} + \Psi.
	\end{split}
\end{equation} To estimate this model, we use the R package {\tt bfa} from \cite{murray2013bayesian}.  The software allows us to sample the covariance as well as the mean via a simple Gibbs step assuming a normal prior on $\mu$.  Factor modeling received considerable theoretical treatment and finance application in \cite{lopes2004bayesian}, \cite{carvalho2008high}, and \cite{wang2011dynamic}, among many others.

\paragraph{Results}

\begin{figure}[H]
\centering
  \includegraphics[scale=.4]{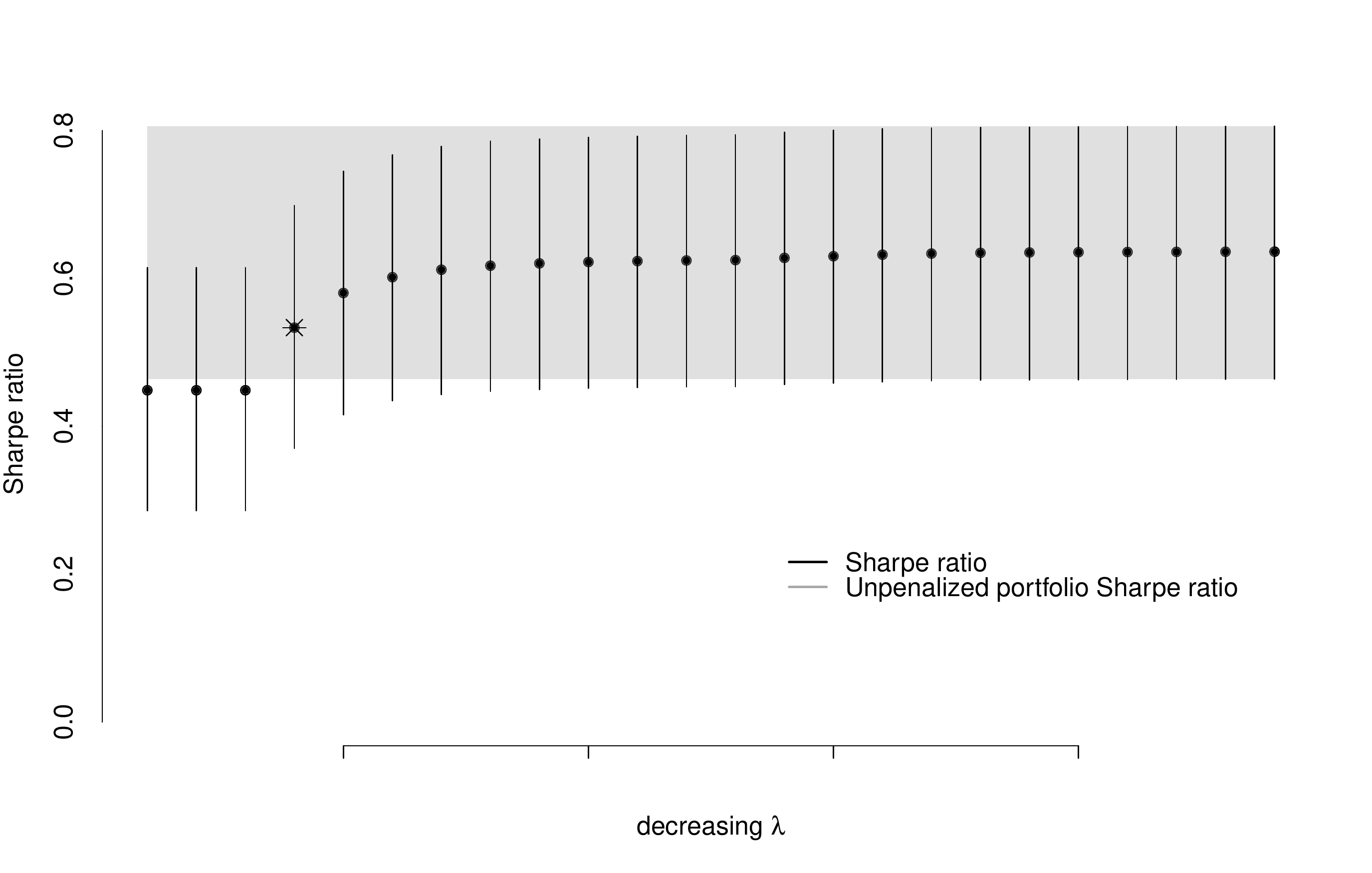}
  \caption{Sharpe ratios of optimal portfolios formed on 25 ETFs.  February 1992 to February 2015. Gray band is the 60\% posterior uncertainty region. Parameters are estimated via the latent factor model.}
  \label{ETFlossgraphnoshortLF}
\end{figure}

After estimating the mean and covariance of 25 ETFs using data from February 1992 to February 2015 under the latent factor model, we display the solution path for the optimization in figure \ref{ETFlossgraphnoshortLF}.  The y-axis displays the Sharpe ratio, and the x-axis are $\lambda$'s from  small (sparse portfolios) to large (full portfolios investing in all ETFs).  From left to right, the x-axis can also be thought of as displaying decreasing portfolio simplicity.  The gray region shows the posterior uncertainty in the Sharpe ratio for the unpenalized optimal portfolio.  As prescribed in \cite{HahnCarvalho} and detailed in Section 2, we choose the largest $\lambda$ such that we remain within this band of uncertainty.  The selected portfolio's Sharpe ratio is represented by the black star.  Given the large amount of posterior uncertainty in the Sharpe ratio, we are able to select a sparse ETF portfolio with high confidence in its ex ante predictive loss relative to the unpenalized optimal portfolio.  In other words, the large uncertainty in the Sharpe ratio metric admits a simple portfolio choice while simultaneously showing we are probably not giving up a lot.  This uncertainty region can be chosen freely.  For example, one could choose a smaller uncertainty band and trade off between less confidence in the predictive loss, a portfolio with more ETFs, and higher potential Sharpe ratio.

%

Table \ref{ETFtab2LF} shows the selected portfolio and amount invested in each ETF.  The two broad market ETFs are IWR and RSP.  Prevailing wisdom in investing circles says to invest ``in the market," and we see that the chosen portfolio has this profile with additional positions in sector specific ETFs. It is interesting to note the portfolio's exposure to the real estate and technology sectors.  These sectors provide additional diversification (reflected in a higher Sharpe ratio).


\begin{table}[H]
\begin{center}
\footnotesize
\begin{tabular}{|m{1cm}|c|c|c|c|}
\hline
\textbf{ETF} &    IWR  &    IYR  &    IYW  \\ \hline
\textbf{weight} &    23.9 &    14.8 &    61.3\\ \hline
\textbf{style} & \textit{mid-cap} & \textit{real estate} & \textit{tech}  \\ \hline
\end{tabular}
\end{center}
\caption{Selected ETF portfolio from latent factor model.}
\label{ETFtab2LF}
\end{table}

\paragraph{Including a market ETF}

\begin{figure}[H]
\centering
  \includegraphics[scale=.4]{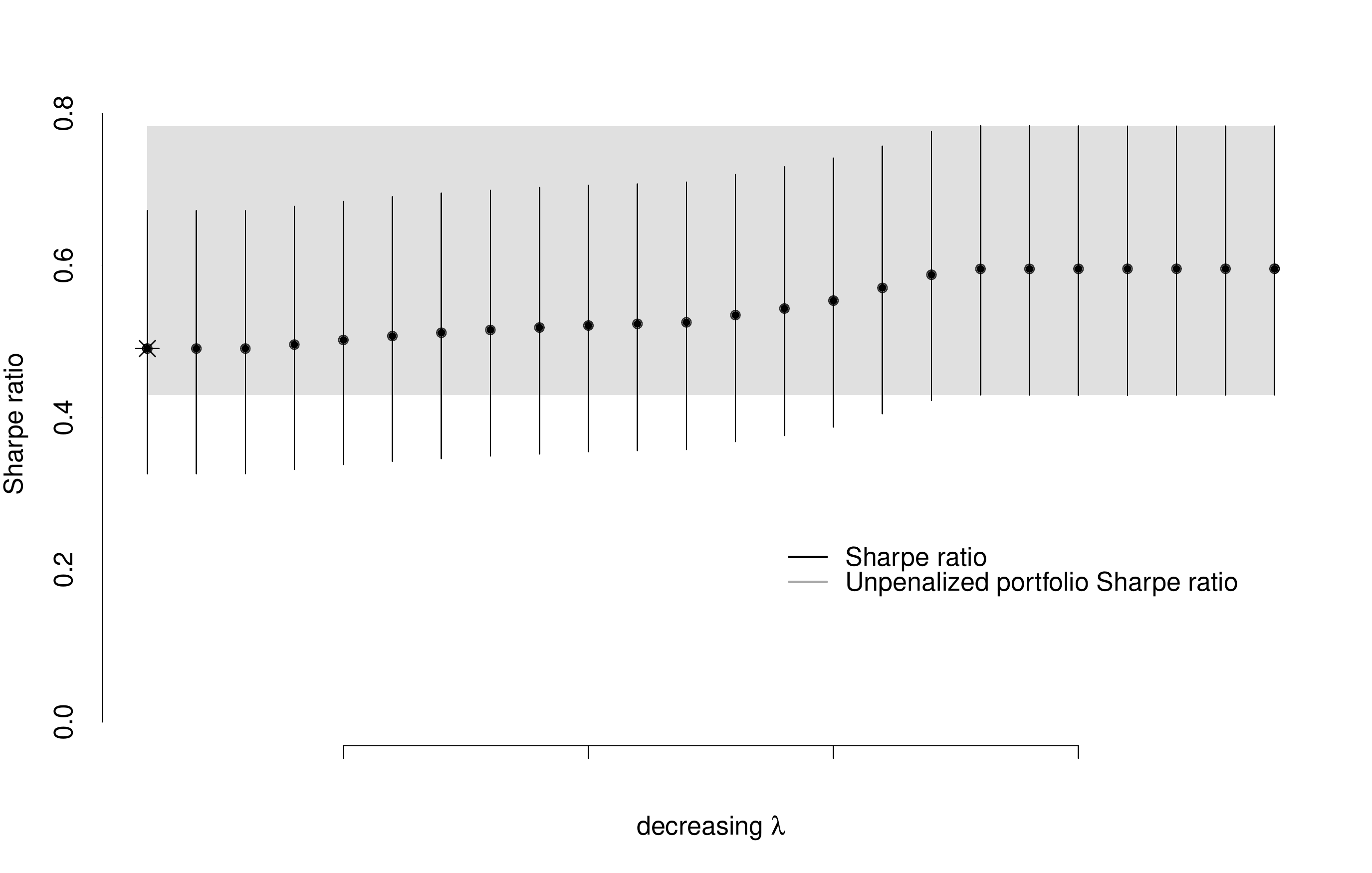}
  \caption{Sharpe ratios of optimal portfolios formed on 25 ETFs.  February 1992 to February 2015. Gray band is the 60\% posterior uncertainty region. Parameters are estimated via the latent factor model.  SPY is unpenalized in the optimization.}
  \label{ETFlossgraphnoshortmarin}
\end{figure}

An individual investor may desire to hold a broad market ETF like SPY, a fund that tracks the S\&P 500 index, with certainty.  This can easily be achieved in the optimization step by unpenalizing SPY. To do this, we fix the penalty parameter associated with SPY to be zero.  As a result, SPY is the first ETF to enter and remains a component of all portfolios along the solution path. Figure \ref{ETFlossgraphnoshortmarin} and table \ref{ETFtab2marin} show the solution path and selected portfolio using the latent factor model and when SPY is unpenalized in the optimization.

The uncertainty in the Sharpe ratio results in selecting the first ETF that appears in the solution path, SPY.  This is an important result of our decision-theoretic framework.  If uncertainty in the Sharpe ratio for the best portfolio (from the unpenalized solution) contains the entire solution path, then not much is lost in predictive deterioration of the Sharpe ratio by choosing the simplest portfolio.  Thus, a long only investor desiring simplicity should invest in only SPY.  This result harkens back to the beginnings of asset pricing theory.  In the mid-1960s, the derivation of the Capital Asset Pricing Model (CAPM) depended crucially on the assumption that the market portfolio is the maximum Sharpe ratio portfolio \citep{Lintner,Sharpe}.  After integrating over parameter and return uncertainty in a simple decision-theoretic framework 50 years later, we have arrived at a similar conclusion:  A market ETF, such as SPY, may be the best solution for a mean-variance investor.

\begin{table}[H]
\begin{center}
\footnotesize
\begin{tabular}{|m{1cm}|c|c|c|c|}
\hline
\textbf{ETF} &    SPY  \\ \hline
\textbf{weight} &   100 \\ \hline
\textbf{style} & \textit{market}  \\ \hline
\end{tabular}
\end{center}
\caption{Selected ETF portfolio from latent factor model.  SPY is unpenalized in the optimization.}
\label{ETFtab2marin}
\end{table}



\subsubsection{Normal-inverse-Wishart model}

\paragraph{Model overview}

In this section, we consider a model with conjugacy for the mean and covariance.  A useful reference detailing this model is \cite{meucci2009risk}.  Asset returns are modeled as:

\begin{equation}
	\begin{split}
		\tilde{R} &\sim \text{N}(\mu,\Sigma)
		\\
		&(\mu,\Sigma) \sim \text{Normal-inverse-Wishart}(\mu_{0},\kappa_{0},\nu_{0},\Sigma_{0}),
	\end{split}
\end{equation}where $\mu$ and $\Sigma$ are the first and second moments of the asset returns, respectively.  Given the Normal-inverse-Wishart prior on the joint distribution and assuming we have $n$ realizations of the asset return vector, $r$, the posterior distribution is:

\begin{equation}
	\begin{split}
		(\mu,\Sigma) \vert r &\sim \text{Normal-inverse-Wishart}\left(\frac{\kappa_{0}\mu_{0} + n\overline{r}}{\kappa_{0} + n}, \kappa_{0} + n, \nu_{0} + n, V \right)
		\\
		V &= \left( v_{0}\Sigma_{0} + n C_{r} + \frac{\kappa_{0}n}{\kappa_{0} + n}(\overline{r} - \mu_{0})(\overline{r} - \mu_{0})^{T}  \right),
	\end{split}
\end{equation} where $C_{r}$ is the sample covariance of the returns.  It is straightforward to sample this posterior by sequentially sampling the marginal distribution $p(\Sigma)$ and conditional distribution $p(\mu \vert \Sigma)$.

\paragraph{Results and comparisons}

\begin{figure}[H]
\centering
  \includegraphics[scale=.4]{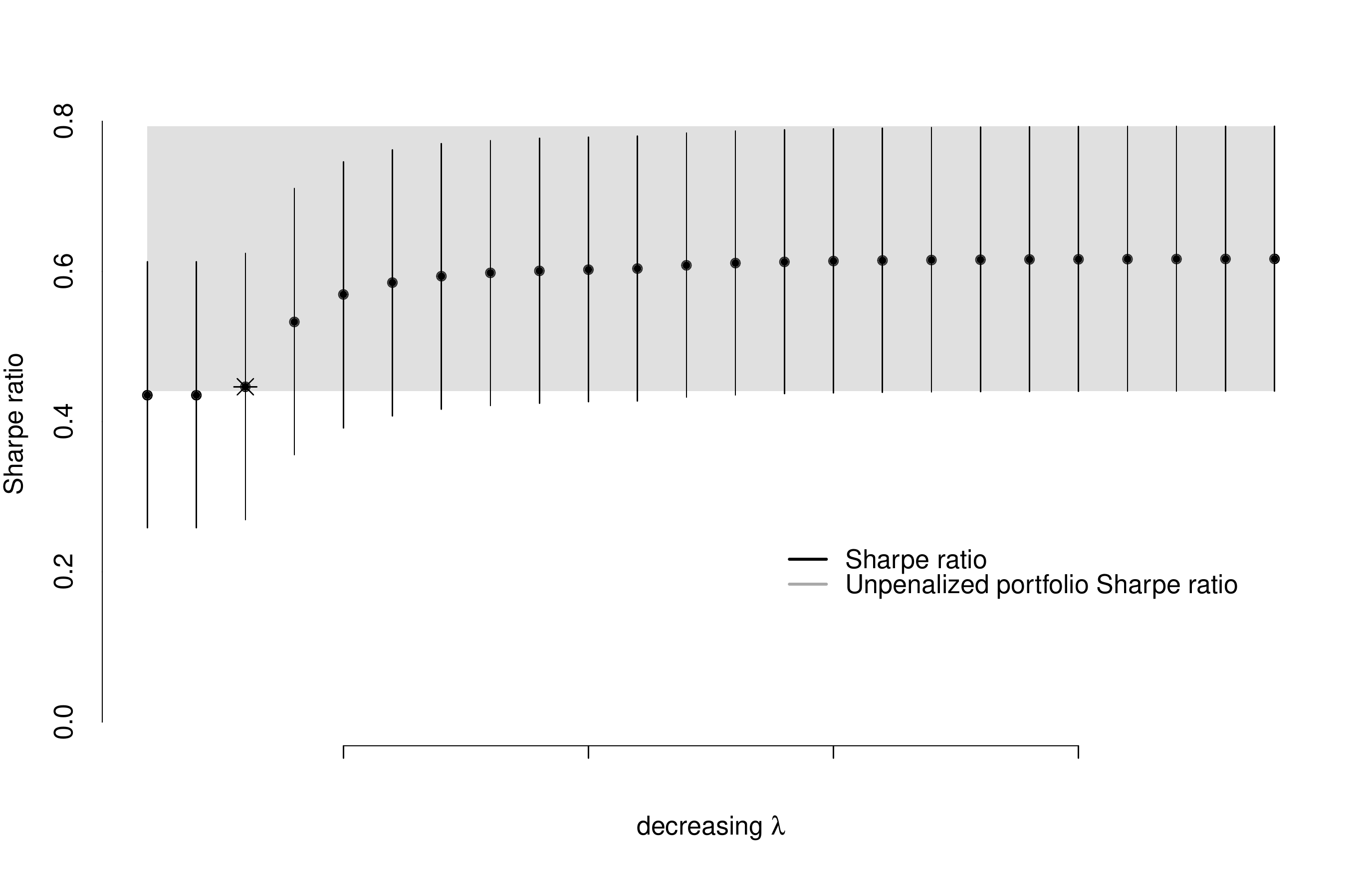}
  \caption{Sharpe ratios of optimal portfolios formed on 25 ETFs.  February 1992 to February 2015. Gray band is the 60\% posterior uncertainty region. Parameters are estimated via the NIW model.}
  \label{ETFlossgraphnoshortniw}
\end{figure}

Here, we show results using the mean and variance parameter estimations from the normal-inverse-Wishart model.  Recall that the prior distribution is also a normal-inverse-Wishart with parameters $(\mu_{0},\kappa_{0},\nu_{0},\Sigma_{0})$.  We choose an uninformative prior for the parameters and perform the same analysis as before.  Figure \ref{ETFlossgraphnoshortniw} shows the solution path for the penalized optimization.  As expected, it is similar to the latent factor model solution path displayed in figure \ref{ETFlossgraphnoshortLF}. However, due to the difference in models, the posterior uncertainty band for the normal-inverse-Wishart model is wider, demonstrating how choice of prior and likelihood enter into the decision making step of portfolio selection. This highlights a key aspect of our methodology: sparsification is influenced by inference through posterior uncertainty. Therefore, the model need not be poorly (or incorrectly) specified by conflating model specification with the investor's preference for sparsity.

%
%
%

The chosen ETF portfolio for the normal-inverse-Wishart model is shown in table \ref{ETFtab2niw}.  It is allocated to the real estate and technology sectors, and the posterior mean of the Sharpe ratio is slightly above 0.4. The choice of such a small portfolio is due to the large uncertainty region of the unpenalized optimal portfolio.  Given the band of posterior uncertainty in figure \ref{ETFlossgraphnoshortniw}, we interpret this Sharpe ratio as follows: With posterior probability 99\%, this Sharpe ratio is no more than $\sim 0.2 \hspace{2mm} (=0.6-0.4)$ away from the unpenalized portfolio's Sharpe ratio, roughly around 0.6.  This is a reasonable difference since the unpenalized portfolio has several other long positions permitting diversification and increases in the Sharpe ratio.

\begin{table}[H]
\begin{center}
\footnotesize
\begin{tabular}{|m{1cm}|c|c|c|}
\hline
\textbf{ETF} &    IYR  &    IYW \\ \hline
\textbf{weight} &    4.7  &    95.3 \\ \hline
\textbf{style} & \textit{real estate} & \textit{tech}  \\ \hline
\end{tabular}
\end{center}
\caption{Selected ETF portfolio from NIW model.}
\label{ETFtab2niw}
\end{table}

\begin{figure}[H]
\centering
  \includegraphics[scale=.4]{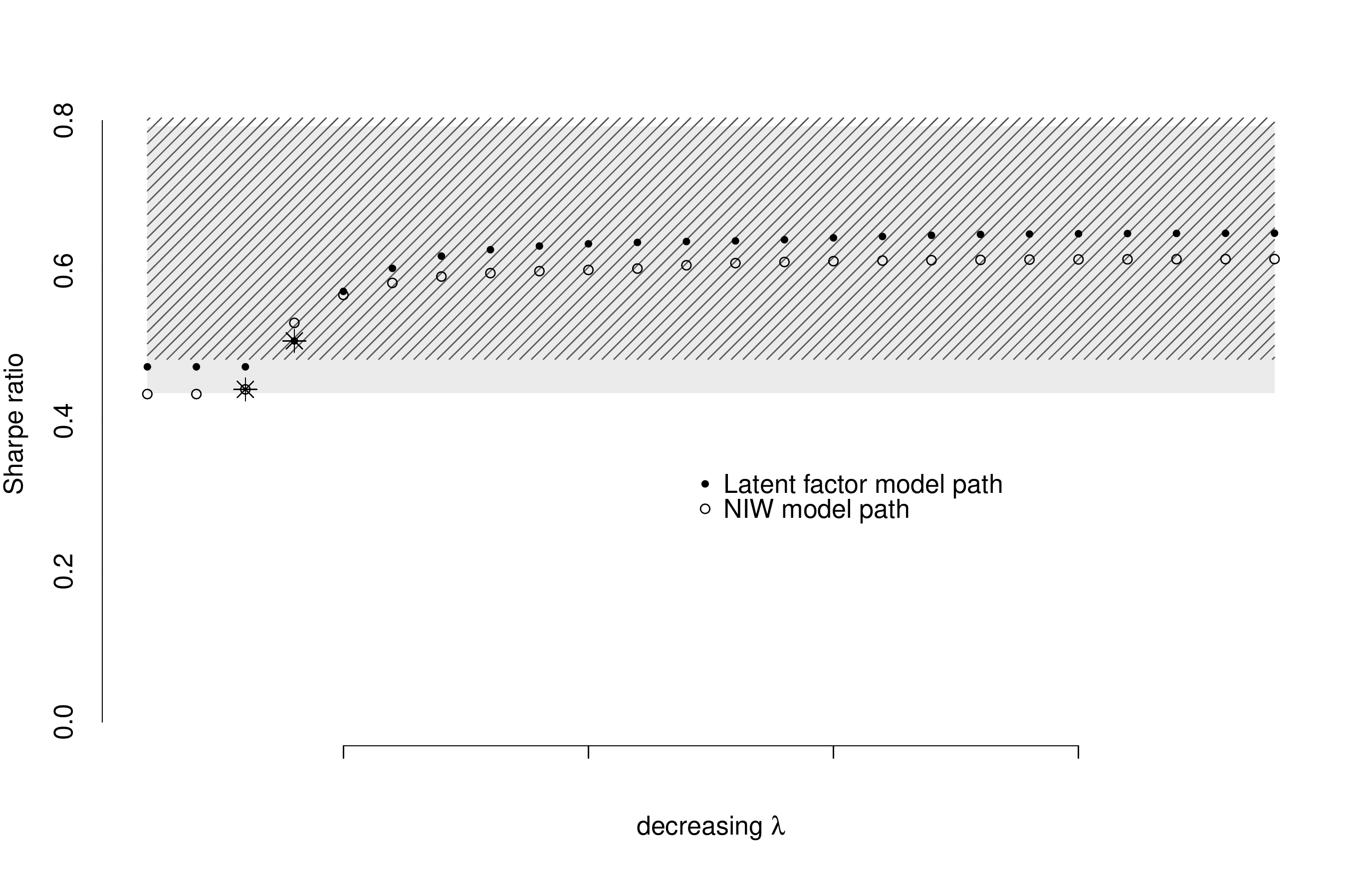}
  \caption{NIW and latent factor model comparison: Sharpe ratios of optimal portfolios formed on 25 ETFs (short selling allowed).  February 1992 to February 2015. Shaded bands are 60\% posterior uncertainty regions for the two models.}
  \label{ETFlossgraphcomparepaths}
\end{figure}

Figure \ref{ETFlossgraphcomparepaths} displays the solution paths from the latent factor and NIW models for direct comparison as well as the unpenalized optimal portfolio uncertainty regions.  The latent factor model region is displayed in diagonal black lines, and the NIW model region is displayed in solid gray.  As previously mentioned, the two solution paths of the Sharpe ratio posterior means as well as the uncertainty regions differ slightly, and this is due to model specification.  This plot emphasizes an equivalence interpretation that can be made when comparing statistical models along the lines of: For a given level of confidence (in our example, 60\%), the NIW model gives a more sparsified portfolio of two ETFs than the latent factor model portfolio of three ETFs.  In other words, our methodology admits a simple way to compare levels of portfolio simplicity among several statistical models for fixed level of posterior confidence.

\subsection{A practical example using a dynamic model} \label{DLMsection}

An investor may desire the flexibility to update their portfolio allocation over time -- particularly when future realizations of asset returns may modify their views of the moments $\mu$ and $\Sigma$.  In this case, we are interested in estimating time-varying moments $\mu_{t}$ and $\Sigma_{t}$ and using them to make a portfolio decision for time $t+1$.  Analogous to the first half of this section, the portfolio decision will rely on the mean-variance objective spelled out in section \ref{LossFunctionDerivation} -- now a function of moments of returns that vary in time.  We assume the investor is myopic and uses moment estimates dependent on all available data to construct a portfolio for the next time period, thus solving the following optimization problem:

\begin{equation}\label{optprobfinaltimevary}
	\begin{split} 
		 \min_{w_{t+1}\geq0} \hspace{3mm} \frac{1}{2} \norm{ L_{t}^{T}w_{t+1}-L_{t}^{-1}\overline{\mu_{t}} }_{2}^{2} + \lambda \norm{ w_{t+1} }_{1}.
	\end{split}
\end{equation}where $\overline{\mu_{t}}$ and $\overline{\Sigma_{t}} = L_{t}L_{t}^{T}$ are the posterior means of the time-varying moments of asset returns. Optimization \ref{optprobfinaltimevary} is the time dependent version of the original optimization \ref{optprobfinal}.  This comprises an out-of-sample analysis in that only available data is used to make a decision at an uncertain next time step.

First, we will describe the model for $\mu_{t}$ and $\Sigma_{t}$.  Second, we will show how these moments may be used in a dynamic ETF investing strategy.

\paragraph{Model overview}

The dynamic regression model presented is related to the recently proposed five factor linear model of \cite{FF5}.  This model states that asset returns are linearly proportional to a set of five asset pricing factors which are themselves tradable portfolios.  Specifically, we model the joint distribution of future ETF and asset pricing factor returns $p(\tilde{R}_{t},\tilde{R}_{t}^{F})$ compositionally by modeling $p(\tilde{R}_{t} \mid \tilde{R}_{t}^{F})$ and $p(\tilde{R}_{t}^{F})$.  Following the dynamic linear model setup from \cite{harrison1999bayesian}, we model the future return of ETF $i$ as a linear combination of future factor returns ($\tilde{R}_{t}^{i} \mid \tilde{R}_{t}^{F}$):



\begin{equation}\label{DLM1}
	\begin{split}
		\tilde{R}_{t}^{i} &= \beta_{t}^{i}\tilde{R}_{t}^{m} + \epsilon_{t}^{i}, \hspace{6mm} \epsilon_{t}^{i} \sim N(0,1/\phi_{t}^{i}),
		\\
		\beta_{t}^{i} &= \beta_{t-1}^{i} + w_{t}^{i}, \hspace{6mm} w_{t}^{i} \sim \text{T}_{n_{t-1}^{i}}(0,W_{t}^{i}),
		\\
		\vspace{0mm}
		\\
		\beta_{0}^{i} &\mid D_{0} \sim \text{T}_{n_{0}^{i}}(m_{0}^{i},C_{0}^{i}),
		\\
		\phi_{0}^{i} &\mid D_{0} \sim \text{Ga}(n_{0}^{i}/2,d_{0}^{i}/2),
		\\
		\vspace{0mm}
		\\
		\beta_{t}^{i} &\mid D_{t-1} \sim \text{T}_{n_{t-1}^{i}}(m_{t-1}^{i},R_{t}^{i}), \hspace{6mm} R_{t}^{i} = C_{t-1}^{i}/\delta_{\beta},
		\\
		\phi_{t}^{i} &\mid D_{t-1} \sim \text{Ga}(\delta_{\epsilon} n_{t-1}^{i}/2,\delta_{\epsilon} d_{t-1}^{i}/2),
	\end{split}
\end{equation}where $W_{t}^{i} = \frac{1-\delta_{\beta}}{\delta_{\beta}}C_{t-1}^{i}$. This model permits the coefficients on the factors as well as the observation and state level variances to vary in time.  Pre-specified discount factors $\delta_{\epsilon}$ and $\delta_{\beta}$ $\in (0,1)$ accomplish this goal for the observation and state level variances, respectively.  

We model the five factor future returns $\tilde{R}_{t}^{F}$ with a full residual covariance matrix using the following matrix normal dynamic linear model:




\begin{equation}\label{DLM2}
	\begin{split}
		\tilde{R}_{t}^{F} &= \mu_{t}^{F} + \nu_{t} \hspace{10mm} \nu_{t} \sim \text{N}(0,\Sigma_{t}^{F}), 
		\\
		\mu_{t}^{F} &= \mu_{t-1}^{F} + \Omega_{t} \hspace{6mm} \Omega_{t} \sim \text{Matrix Normal}(0,W_{t},\Sigma_{t}^{F}),
		\\
		\vspace{0mm}
		\\
		(\mu_{0}^{F}&,\Sigma_{0}^{F} \mid D_{0}) \sim \text{NW}_{n_{0}}^{-1}(m_{0},C_{0},S_{0}),
		\\
		\vspace{0mm}
		\\
		( \mu_{t}^{F}&, \Sigma_{t}^{F} \mid D_{t-1}) \sim \text{NW}_{\delta_{F}n_{t-1}}^{-1}(m_{t-1},R_{t},S_{t-1}),\hspace{6mm} R_{t} = C_{t-1}/\delta_{c},
	\end{split}
\end{equation}where $W_{t} = \frac{1-\delta_{c}}{\delta_{c}}C_{t-1}$.  Analogous to model \ref{DLM1}, the discount factors $\delta_{F}$ and $\delta_{c}$ in model \ref{DLM2} serve the same purpose of permitting time variation in the observation and state level variances, respectively. An added benefit of \ref{DLM2} is that $\Sigma_{t}^{F}$ is a full residual covariance matrix.  

Models \ref{DLM1} and \ref{DLM2} together constitute a time-varying model for the joint distribution of future ETF and factor returns: $p(\tilde{R}_{t},\tilde{R}_{t}^{F}) = p(\tilde{R}_{t} \mid \tilde{R}_{t}^{F})p(\tilde{R}_{t}^{F})$. As detailed in \cite{harrison1999bayesian}, they are Bayesian models that have closed form posterior distributions of all parameters at each time $t$, and the absence of MCMC is convenient for fast updating and uncertainty characterization -- a necessary ingredient for our portfolio selection procedure.  Under these models, we obtain the following mean and covariance structure for the ETF returns.

\begin{equation}\label{moments}
	\begin{split}
		\mu_{t} &= \beta_{t}^{T}\mu_{t}^{F}
		\\
		\Sigma_{t} &= \beta_{t} \Sigma_{t}^{F} \beta_{t}^{T} + \Psi_{t}
	\end{split}
\end{equation}where column $i$ of $\beta_{t}$ are the coefficients on the factors for ETF $i$, $\beta_{t}^{i}$.  Also, $\Psi_{t}$ is a diagonal matrix with $i$th element $\Psi_{tii} = 1/\phi_{t}^{i}$.  

Posterior means of moments \ref{moments} are obtained by filtering estimations of models \ref{DLM1} and \ref{DLM2} and may be substituted into optimization problem \ref{optprobfinaltimevary} to solve the portfolio selection problem at each time step.  Similar to the static portfolio case, we solve the optimization problem (now \textit{at each time $t$}) along the solution path and choose the simplest portfolio that remains within the uncertainty region for the Sharpe ratio of the unpenalized portfolio.  This procedure is done at each time $t$ to provide a simple ETF portfolio that updates over time.

\paragraph{Results}

In the analysis below, we use return data on 25 ETFs from February 1992 to February 2015.  The returns on the Fama-French five factors are obtained from the publicly available data library on Ken French's website\footnote{http://mba.tuck.dartmouth.edu/pages/faculty/ken.french/}.  We start with 36 months of data to train the model and begin forming portfolios in February 1995. Since models \ref{DLM1} and \ref{DLM2} are flexible in that several discount factors may be specified for the variances, we focus our analysis on a reasonable choice for these discount factors.  Specifically, we fix $\delta_{c} = \delta_{\beta} = 1$ and consider time-varying residual variances $\delta_{F} =\delta_{\epsilon}=0.999$.  Evidence of time-varying residual variance is well-documented in the finance literature (see for example, \cite{ng1991tests}).  

Similar to the static portfolio analysis, we consider forming portfolios that hold a market ETF (SPY) and are long only.  Intuitively, we are interested in constructing a dynamic portfolio strategy for the individual investor that desires market exposure and a couple more diversifying long positions.

\begin{table}[H]
\begin{center}
\footnotesize

\begin{tabular}{|c|c|c|c|c|c|}
  \hline
\textbf{ETF} & SPY & EEM & IWD & IWB & IYR \\
\hline 
 \textbf{style} & market & EM & large value & large & real estate \\
  \hline
199502 & 97.9 & 2.13 & - & - & - \\ 
  199602 & 100 & - & - & - & - \\ 
  199702 & 100 & - & - & - & - \\ 
  199802 & 59 & - & - & - & 41 \\ 
  199902 & 100 & - & - & - & - \\ 
  200002 & 100 & - & - & - & - \\ 
  200102 & 99 & - & - & - & 1 \\ 
  200202 & 42.4 & - & - & - & 57.6 \\ 
  200302 & 9.17 & - & 18.2 & 18.1 & 54.5 \\ 
  200402 & 9.54 & - & 15.7 & 19.8 & 54.9 \\ 
  200502 & 7.61 & - & 20.9 & 26.8 & 44.7 \\ 
  200602 & 14.1 & - & 42.2 & - & 43.7 \\ 
  200702 & 12.3 & - & 46.8 & - & 40.9 \\ 
  200802 & 54.4 & - & - & - & 45.6 \\ 
  200902 & 71.6 & - & - & - & 28.4 \\ 
  201002 & 80.7 & - & - & - & 19.3 \\ 
  201102 & 100 & - & - & - & - \\ 
  201202 & 100 & - & - & - & - \\ 
  201302 & 100 & - & - & - & - \\ 
  201402 & 100 & - & - & - & - \\ 
  201502 & 100 & - & - & - & - \\ 
   \hline
\end{tabular}

\end{center}
\caption{Sparse ETF portfolio weights for the dynamic linear model with $\delta_{F}= \delta_{\epsilon}= 0.999$.  Note that annual weights are shown for brevity although portfolios are updated monthly.}
\label{DLMweights}
\end{table}

Table \ref{DLMweights} shows the weights for each year of the dynamic portfolio strategy.  Note that the portfolios are updated monthly as new data is observed -- annual weights are only shown for brevity. Throughout the trading period, SPY is always included in the portfolio since it is unpenalized in the optimization.  It is underweighted from 2003 through 2008 and is fully allocated to after the financial crisis (2009 to 2015).  Diversifying positions include emerging market stocks at the beginning and large and value stocks and real estate in the middle of the trading period.  The procedure's selection of the real estate ETF and significant allocation leading up to the financial crisis contribute to its performance over the trading period.

\begin{figure}[H]
\centering
  \includegraphics[scale=.43]{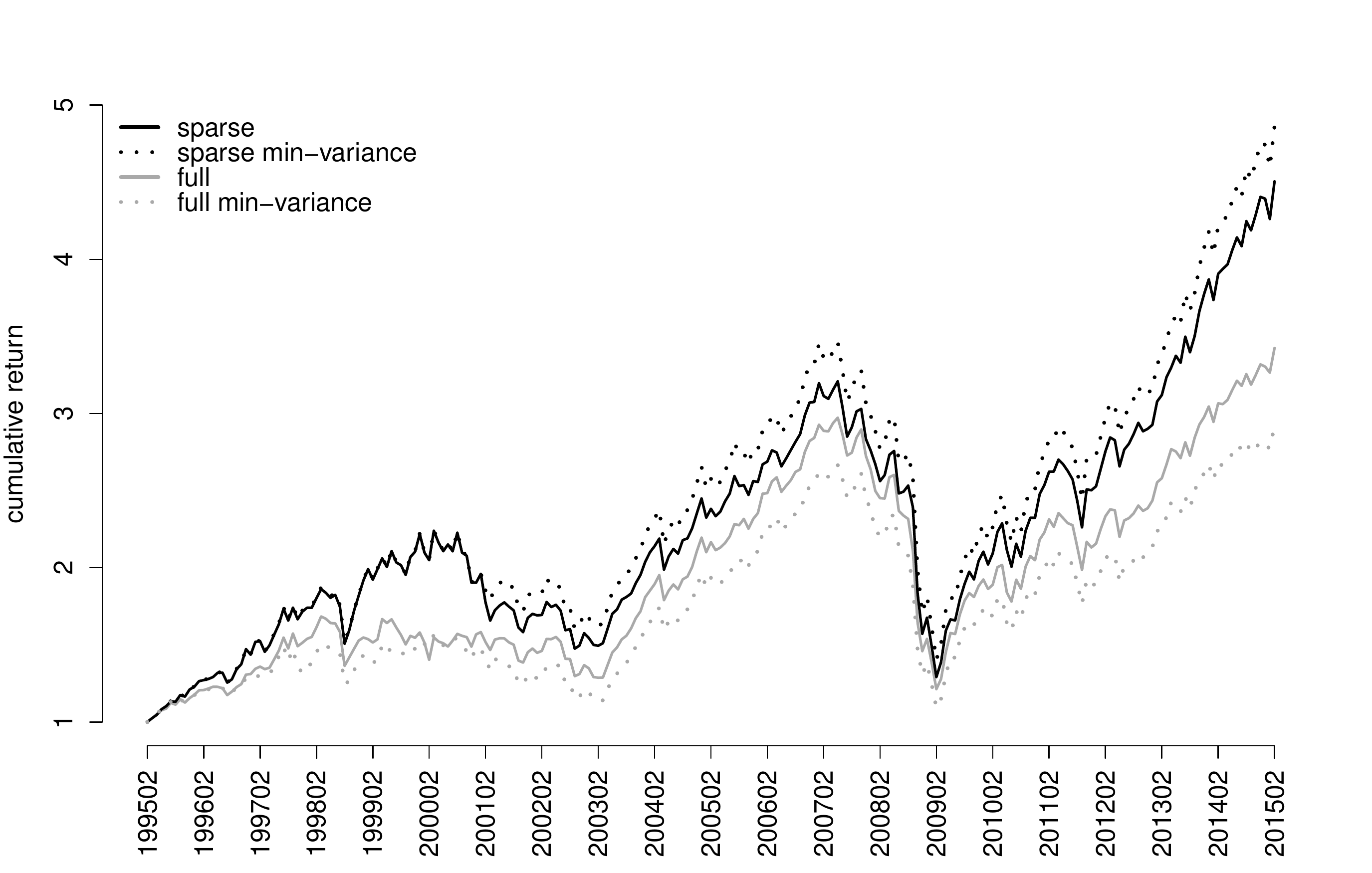}
  \caption{Comparing sparse ETF portfolio out-of-sample performance to their dense (or full) portfolio counterparts.}
\label{OOSgraph1}
\end{figure}

Figure \ref{OOSgraph1} shows the growth of \$1 invested in the dynamic portfolio selection strategy (also know as the cumulative return).  This is shown by the black line and labeled as \textit{sparse}.  Also shown are three other strategies: (\textit{i}) sparse min-variance (dotted black), (\textit{ii}) full (gray), and (\textit{iii}) full min-variance (dotted gray). 

Sparse min-variance is a portfolio formed with the same set of selected ETFs as the sparse portfolio (shown in table \ref{DLMweights}), but the optimal weights are found using a minimum variance objective.  This objective is analogous to mean-variance except the mean vector is replaced with a vector of ones.  For comparison, we also show \textit{full} portfolios that are formed with all 25 ETFs at each point in time.  The key takeaway from this figure is the notable outperformance of the sparse portfolios compared to the full portfolios.  This exhibits an added benefit of implementing a sparse portfolio strategy beyond portfolio simplicity: out-of-sample performance is significantly improved compared to the alternative of optimally investing in all ETFs.

\begin{figure}[H]
\centering
  \includegraphics[scale=.43]{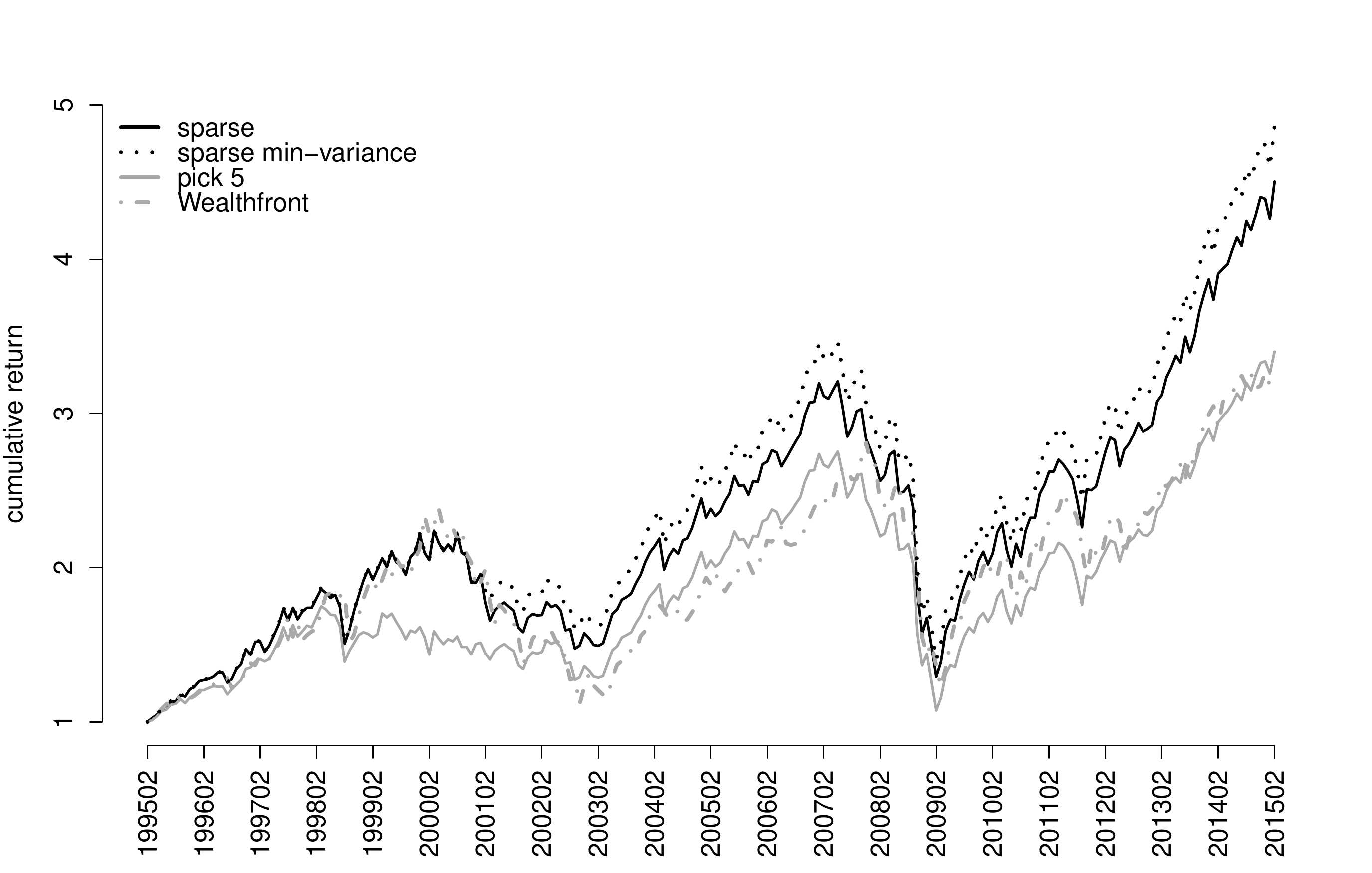}
  \caption{Comparing sparse ETF portfolio out-of-sample performance to two other practical alternatives: (i) A portfolio of the first 5 ETFs along the solution path and, (ii) A model portfolio from the investment manager Wealthfront.com.}
  \label{OOSgraph2}
\end{figure}

Figure \ref{OOSgraph2} compares the sparse portfolio strategies to two other reasonable alternatives: (\textit{i}) pick 5 (gray) and (\textit{ii}) Wealthfront (dotted gray).  The pick 5 portfolio invests in the first five ETFs along the solution path at each point in time.  The underperformance of this portfolio relative to the sparse portfolios highlights the benefit of selecting a \textit{different number} of ETFs depending on the chosen posterior uncertainty region. There appear to be out-of-sample gains from permitting flexibility in the number ETF positions. The Wealthfront portfolio is a strategic portfolio from ETF investment advisor Wealthfront.com. Its positions are: 50\% SPY (market), 15\% EEM (emerging markets), 30\% EZU (eurozone), and 5\% IYW (tech). This fixed ETF portfolio also underperforms the sparse portfolios whose positions and ETF positions change over time.

Table \ref{OOStable} compares the out-of-sample Sharpe ratio and standard deviation of returns for all seven strategies considered.  The sparse strategies shown at the top should be contrasted with the full portfolio strategies (investing in all available ETFs) and other practical strategies shown in the middle and last, respectively.  The Sharpe ratios for the sparse portfolios outperform each of the other five, and the min-variance strategy possesses the best risk-return tradeoff.  Also, the standard deviation for the sparse strategies is slightly higher than the others since these portfolios generally have fewer positions.  However, this added out-of-sample risk is made up for by increased return.

\begin{table}[H]
\begin{center}
\footnotesize

\begin{tabular}{l*2{>{\centering\arraybackslash}m{0.7in}} @{}m{0pt}@{}}
 & \multicolumn{2}{c}{out-of-sample statistics} \\[1mm]
 \cline{2-3}
 &  Sharpe ratio & s.d. &\\[1ex] 
  \hline
\textbf{sparse} & 0.56 & 15.74 \\ 
\textbf{sparse min-var} & 0.59 & 15.53 \\ 
  \hline
  \textbf{full} & 0.50 & 14.49 \\ 
  \textbf{full min-var} & 0.44 & 14.46 \\ 
   \hline
  \textbf{pick 5} & 0.48 & 15.09 \\ 
  \textbf{Wealthfront} & 0.44 & 17.18  \\ 
  \textbf{SPY} & 0.52 & 15.15 \\ 
  \hline
\end{tabular}

\end{center}
\caption{Comparison of out-of-sample portfolio statistics for the seven portfolio strategies considered.  The sparse ETF portfolios notably outperform the dense (or full) portfolios as well as the two practical alternatives and investing in the market ETF, SPY.}
\label{OOStable}
\end{table}

\section{Discussion}

This paper presents a new approach to mean-variance portfolio optimization from the investor's perspective.  A loss function based on the Markowitz objective function is used in tandem with a new procedure that explicitly separates statistical inference from selection of the final portfolio.  We illustrate the procedure by comparing two static models of asset returns and propose an out-of-sample ETF portfolio selection strategy based on a dynamic model for asset returns.

\subsection{Alternative utility specifications}

We end this paper by briefly discussing extensions of the presented methodology.  The flexibility of the portfolio selection procedure to different utility specifications is an important characteristic of this paper. Two relevant examples were mentioned by the referees: (\textit{i}) incorporation of fees and (\textit{ii}) minimization of transaction costs.  In each case, a variant of loss function \ref{optprobfinal2} may be considered.  Fees of the funds can be incorporated directly into the first moment of returns.  For example, suppose a vector of expense ratios (percentage fee charged of total assets managed) of all funds were given by $\tau$.  The first moment of returns within the loss function may be adjusted by $\tau$ to reflect an investor's sensitivity to fees:  

\begin{equation}
	\begin{split} \label{optprobconclusion}
		 \mathcal{L}(w) = \frac{1}{2} \norm{ L^{T}w-L^{-1}(\overline{\mu} -\tau) }_{2}^{2} + \lambda \norm{ w }_{1},
	\end{split}
\end{equation}where $\overline{\mu} -\tau$ is the net return on investment in each fund. The expense ratio can also be incorporated at various time frequencies depending the units of $\overline{\mu}$.  For example, if $\overline{\mu}$ represents an expected monthly fund return, then $\tau$ (typically quoted annually), can be expressed on a monthly scale, too. 

Sensitivity to transaction costs can be similarly accounted for by modifying the dynamic version of the mean-variance loss function (the objective in \ref{optprobfinaltimevary}).  This can be accomplished by penalizing the difference in consecutive weight vectors through time, $w_{t+1} - w_{t}$.  An example loss function would look like:
\begin{equation}
	\begin{split} \label{optprobconclusion2}
		 \mathcal{L}(w_{t+1}) = \frac{1}{2} \norm{ L_{t}^{T}w_{t+1}-L_{t}^{-1}\overline{\mu_{t}} }_{2}^{2} + \lambda \norm{ w_{t+1} - w_{t} }_{1}.
	\end{split}
\end{equation}This penalty is designed to encourage slow movement in portfolio positions (given by the weight vector) over time so as to avoid frequent buying and selling of assets.  In our empirical results, we focus on the original loss function in optimization \ref{optprobfinaltimevary} for two reasons.  First, ETFs are low-cost investments and $\tau$ is small relative to $\overline{\mu}$.  Second, our sparse dynamic portfolio weights (shown in table \ref{DLMweights}) under a variety of modeling choices vary gradually without a penalty on the weight difference.

\newpage
\appendix

\section{Regularized portfolio optimization}

Regularization methods for portfolio optimization follow two schools of thought.  The first attacks estimation errors directly by attempting to regularize statistical estimates.  This is often done in a Bayesian context by shrinking the mean estimates to a ``grand mean" \citep{james1961estimation}.  Similar applications of shrinkage used for regularization include \cite{jorion1985international} and \cite{jorion1986bayes} who set the mean return of the minimum variance portfolio to be the grand mean.  In addition to the mean, others have focused on shrinking the covariance matrix in the portfolio selection problem, including \cite{ledoit2003improved}, \cite{ledoit2003honey}, and \cite{garlappi2007portfolio}.  In \cite{garlappi2007portfolio}, they consider parameter and model uncertainty to improve portfolio selection and out-of-sample Sharpe ratio. The second stream of literature focuses on regularizing the portfolio weights directly.  This literature is distinct in that regularization is delegated to the portfolio optimization step in contrast to inference of the optimization inputs.  This is a popular approach since mean-variance optimization can be formulated in terms of a Least Absolute Shrinkage and Selection Operator (LASSO) objective function from \cite{Tib} with fast computation of the optimal solution.  In this setting, the weights in the objective function are penalized with an $l_1$ norm.  Among the first to investigate weight regularization in the context of portfolio optimization is \cite{brodie2009sparse}.  They document improved out-of-sample performance as measured by the Sharpe ratio and consistently beat a naive strategy of investing an equal amount in each asset.  \cite{demiguel2009generalized} is a separate study showing similar results.  These more recent studies have clarified and expanded on earlier work by \cite{jagannathan2002risk} who showed that constraining the weights is equivalent to shrinking sample estimation of the covariance matrix.  \cite{fan2012vast} show that $l_1$ regularization limits gross portfolio exposure, leads to no accumulation of estimation errors, and results in outperformance compared to standard mean-variance portfolios.  Specifically, \cite{fan2012vast} and references therein discuss how to use a constraint on portfolio size to ``bridge the gap" between a no short sale optimization and unconstrained optimization. This paper represents useful theoretical treatment of the ideas presented in \cite{jagannathan2002risk} and their connections with the canonical mean-variance problem.   Recent work on regularization for portfolio selection includes \cite{yen2013note}, \cite{yen2014solving}, \cite{carrasco2011optimal}, \cite{fernandes2012regularized}, \cite{fastrich2013constructing} with applications to index tracking in \cite{fastrich2014cardinality}, \cite{takeda2013simultaneous}, \cite{Wu2}, and \cite{Wu}.


\newpage
\bibliographystyle{imsart-nameyear.bst}
\bibliography{DSSPortfolioOptimization.bib}

\begin{thebibliography}{56}

\bibitem[\protect\citeauthoryear{Barber and Odean}{2000}]{barber2000trading}
\begin{barticle}[author]
\bauthor{\bsnm{Barber},~\bfnm{Brad~M}\binits{B.~M.}} \AND
  \bauthor{\bsnm{Odean},~\bfnm{Terrance}\binits{T.}}
(\byear{2000}).
\btitle{Trading is hazardous to your wealth: The common stock investment
  performance of individual investors}.
\bjournal{Journal of Finance}
\bpages{773--806}.
\end{barticle}
\endbibitem

\bibitem[\protect\citeauthoryear{Barber et~al.}{2009}]{barber2009just}
\begin{barticle}[author]
\bauthor{\bsnm{Barber},~\bfnm{Brad~M}\binits{B.~M.}},
  \bauthor{\bsnm{Lee},~\bfnm{Yi-Tsung}\binits{Y.-T.}},
  \bauthor{\bsnm{Liu},~\bfnm{Yu-Jane}\binits{Y.-J.}} \AND
  \bauthor{\bsnm{Odean},~\bfnm{Terrance}\binits{T.}}
(\byear{2009}).
\btitle{Just how much do individual investors lose by trading?}
\bjournal{Review of Financial studies}
\bvolume{22}
\bpages{609--632}.
\end{barticle}
\endbibitem

\bibitem[\protect\citeauthoryear{Best and Grauer}{1991}]{best1991sensitivity}
\begin{barticle}[author]
\bauthor{\bsnm{Best},~\bfnm{Michael~J}\binits{M.~J.}} \AND
  \bauthor{\bsnm{Grauer},~\bfnm{Robert~R}\binits{R.~R.}}
(\byear{1991}).
\btitle{On the sensitivity of mean-variance-efficient portfolios to changes in
  asset means: some analytical and computational results}.
\bjournal{Review of Financial Studies}
\bvolume{4}
\bpages{315--342}.
\end{barticle}
\endbibitem

\bibitem[\protect\citeauthoryear{Britten-Jones}{1999}]{britten1999sampling}
\begin{barticle}[author]
\bauthor{\bsnm{Britten-Jones},~\bfnm{Mark}\binits{M.}}
(\byear{1999}).
\btitle{The Sampling Error in Estimates of Mean-Variance Efficient Portfolio
  Weights}.
\bjournal{The Journal of Finance}
\bvolume{54}
\bpages{655--671}.
\end{barticle}
\endbibitem

\bibitem[\protect\citeauthoryear{Broadie}{1993}]{broadie1993computing}
\begin{barticle}[author]
\bauthor{\bsnm{Broadie},~\bfnm{Mark}\binits{M.}}
(\byear{1993}).
\btitle{Computing efficient frontiers using estimated parameters}.
\bjournal{Annals of Operations Research}
\bvolume{45}
\bpages{21--58}.
\end{barticle}
\endbibitem

\bibitem[\protect\citeauthoryear{Brodie et~al.}{2009}]{brodie2009sparse}
\begin{barticle}[author]
\bauthor{\bsnm{Brodie},~\bfnm{Joshua}\binits{J.}},
  \bauthor{\bsnm{Daubechies},~\bfnm{Ingrid}\binits{I.}},
  \bauthor{\bsnm{De~Mol},~\bfnm{Christine}\binits{C.}},
  \bauthor{\bsnm{Giannone},~\bfnm{Domenico}\binits{D.}} \AND
  \bauthor{\bsnm{Loris},~\bfnm{Ignace}\binits{I.}}
(\byear{2009}).
\btitle{Sparse and stable Markowitz portfolios}.
\bjournal{Proceedings of the National Academy of Sciences}
\bvolume{106}
\bpages{12267--12272}.
\end{barticle}
\endbibitem

\bibitem[\protect\citeauthoryear{Carrasco and
  Noumon}{2011}]{carrasco2011optimal}
\begin{btechreport}[author]
\bauthor{\bsnm{Carrasco},~\bfnm{Marine}\binits{M.}} \AND
  \bauthor{\bsnm{Noumon},~\bfnm{N{\'e}r{\'e}e}\binits{N.}}
(\byear{2011}).
\btitle{Optimal portfolio selection using regularization}
\btype{Technical Report},
\bpublisher{Discussion paper}.
\end{btechreport}
\endbibitem

\bibitem[\protect\citeauthoryear{Carvalho et~al.}{2008}]{carvalho2008high}
\begin{barticle}[author]
\bauthor{\bsnm{Carvalho},~\bfnm{Carlos~M}\binits{C.~M.}},
  \bauthor{\bsnm{Chang},~\bfnm{Jeffrey}\binits{J.}},
  \bauthor{\bsnm{Lucas},~\bfnm{Joseph~E}\binits{J.~E.}},
  \bauthor{\bsnm{Nevins},~\bfnm{Joseph~R}\binits{J.~R.}},
  \bauthor{\bsnm{Wang},~\bfnm{Quanli}\binits{Q.}} \AND
  \bauthor{\bsnm{West},~\bfnm{Mike}\binits{M.}}
(\byear{2008}).
\btitle{High-dimensional sparse factor modeling: applications in gene
  expression genomics}.
\bjournal{Journal of the American Statistical Association}
\bvolume{103}.
\end{barticle}
\endbibitem

\bibitem[\protect\citeauthoryear{CRSP}{1992-2015}]{CRSP}
\begin{bmisc}[author]
\bauthor{\bsnm{CRSP}}
(\byear{1992-2015}).
\btitle{The Center for Research in Security Prices}.
\bhowpublished{Wharton Research Data Services}.
\end{bmisc}
\endbibitem

\bibitem[\protect\citeauthoryear{DeMiguel, Garlappi and
  Uppal}{2009}]{demiguel2009optimal}
\begin{barticle}[author]
\bauthor{\bsnm{DeMiguel},~\bfnm{Victor}\binits{V.}},
  \bauthor{\bsnm{Garlappi},~\bfnm{Lorenzo}\binits{L.}} \AND
  \bauthor{\bsnm{Uppal},~\bfnm{Raman}\binits{R.}}
(\byear{2009}).
\btitle{Optimal versus naive diversification: How inefficient is the 1/N
  portfolio strategy?}
\bjournal{Review of Financial Studies}
\bvolume{22}
\bpages{1915--1953}.
\end{barticle}
\endbibitem

\bibitem[\protect\citeauthoryear{DeMiguel
  et~al.}{2009}]{demiguel2009generalized}
\begin{barticle}[author]
\bauthor{\bsnm{DeMiguel},~\bfnm{Victor}\binits{V.}},
  \bauthor{\bsnm{Garlappi},~\bfnm{Lorenzo}\binits{L.}},
  \bauthor{\bsnm{Nogales},~\bfnm{Francisco~J}\binits{F.~J.}} \AND
  \bauthor{\bsnm{Uppal},~\bfnm{Raman}\binits{R.}}
(\byear{2009}).
\btitle{A generalized approach to portfolio optimization: Improving performance
  by constraining portfolio norms}.
\bjournal{Management Science}
\bvolume{55}
\bpages{798--812}.
\end{barticle}
\endbibitem

\bibitem[\protect\citeauthoryear{Dickinson}{1974}]{dickinson1974reliability}
\begin{barticle}[author]
\bauthor{\bsnm{Dickinson},~\bfnm{John~P}\binits{J.~P.}}
(\byear{1974}).
\btitle{The reliability of estimation procedures in portfolio analysis}.
\bjournal{Journal of Financial and Quantitative Analysis}
\bvolume{9}
\bpages{447--462}.
\end{barticle}
\endbibitem

\bibitem[\protect\citeauthoryear{Efron et~al.}{2004}]{Efron}
\begin{barticle}[author]
\bauthor{\bsnm{Efron},~\bfnm{Bradley}\binits{B.}},
  \bauthor{\bsnm{Hastie},~\bfnm{Trevor}\binits{T.}},
  \bauthor{\bsnm{Johnstone},~\bfnm{Iain}\binits{I.}},
  \bauthor{\bsnm{Tibshirani},~\bfnm{Robert}\binits{R.}} \betal{et~al.}
(\byear{2004}).
\btitle{Least angle regression}.
\bjournal{The Annals of statistics}
\bvolume{32}
\bpages{407--499}.
\end{barticle}
\endbibitem

\bibitem[\protect\citeauthoryear{Fama and French}{2015}]{FF5}
\begin{barticle}[author]
\bauthor{\bsnm{Fama},~\bfnm{Eugene~F}\binits{E.~F.}} \AND
  \bauthor{\bsnm{French},~\bfnm{Kenneth~R}\binits{K.~R.}}
(\byear{2015}).
\btitle{A five-factor asset pricing model}.
\bjournal{Journal of Financial Economics}
\bvolume{116}
\bpages{1--22}.
\end{barticle}
\endbibitem

\bibitem[\protect\citeauthoryear{Fan, Zhang and Yu}{2012}]{fan2012vast}
\begin{barticle}[author]
\bauthor{\bsnm{Fan},~\bfnm{Jianqing}\binits{J.}},
  \bauthor{\bsnm{Zhang},~\bfnm{Jingjin}\binits{J.}} \AND
  \bauthor{\bsnm{Yu},~\bfnm{Ke}\binits{K.}}
(\byear{2012}).
\btitle{Vast portfolio selection with gross-exposure constraints}.
\bjournal{Journal of the American Statistical Association}
\bvolume{107}
\bpages{592--606}.
\end{barticle}
\endbibitem

\bibitem[\protect\citeauthoryear{Fastrich, Paterlini and
  Winker}{2013}]{fastrich2013constructing}
\begin{barticle}[author]
\bauthor{\bsnm{Fastrich},~\bfnm{Bj{\"o}rn}\binits{B.}},
  \bauthor{\bsnm{Paterlini},~\bfnm{Sandra}\binits{S.}} \AND
  \bauthor{\bsnm{Winker},~\bfnm{Peter}\binits{P.}}
(\byear{2013}).
\btitle{Constructing optimal sparse portfolios using regularization methods}.
\bjournal{Computational Management Science}
\bpages{1--18}.
\end{barticle}
\endbibitem

\bibitem[\protect\citeauthoryear{Fastrich, Paterlini and
  Winker}{2014}]{fastrich2014cardinality}
\begin{barticle}[author]
\bauthor{\bsnm{Fastrich},~\bfnm{Bj{\"o}rn}\binits{B.}},
  \bauthor{\bsnm{Paterlini},~\bfnm{Sandra}\binits{S.}} \AND
  \bauthor{\bsnm{Winker},~\bfnm{Peter}\binits{P.}}
(\byear{2014}).
\btitle{Cardinality versus q-norm constraints for index tracking}.
\bjournal{Quantitative Finance}
\bvolume{14}
\bpages{2019--2032}.
\end{barticle}
\endbibitem

\bibitem[\protect\citeauthoryear{Fernandes, Rocha and
  Souza}{2012}]{fernandes2012regularized}
\begin{bmisc}[author]
\bauthor{\bsnm{Fernandes},~\bfnm{Marcelo}\binits{M.}},
  \bauthor{\bsnm{Rocha},~\bfnm{Guilherme}\binits{G.}} \AND
  \bauthor{\bsnm{Souza},~\bfnm{Thiago}\binits{T.}}
(\byear{2012}).
\btitle{Regularized minimum-variance portfolios using asset group information}.
\end{bmisc}
\endbibitem

\bibitem[\protect\citeauthoryear{Frankfurter, Phillips and
  Seagle}{1971}]{frankfurter1971portfolio}
\begin{barticle}[author]
\bauthor{\bsnm{Frankfurter},~\bfnm{George~M}\binits{G.~M.}},
  \bauthor{\bsnm{Phillips},~\bfnm{Herbert~E}\binits{H.~E.}} \AND
  \bauthor{\bsnm{Seagle},~\bfnm{John~P}\binits{J.~P.}}
(\byear{1971}).
\btitle{Portfolio selection: the effects of uncertain means, variances, and
  covariances}.
\bjournal{Journal of Financial and Quantitative Analysis}
\bvolume{6}
\bpages{1251--1262}.
\end{barticle}
\endbibitem

\bibitem[\protect\citeauthoryear{Frost and Savarino}{1988}]{frost1988better}
\begin{barticle}[author]
\bauthor{\bsnm{Frost},~\bfnm{Peter~A}\binits{P.~A.}} \AND
  \bauthor{\bsnm{Savarino},~\bfnm{James~E}\binits{J.~E.}}
(\byear{1988}).
\btitle{For better performance: Constrain portfolio weights}.
\end{barticle}
\endbibitem

\bibitem[\protect\citeauthoryear{Garlappi, Uppal and
  Wang}{2007}]{garlappi2007portfolio}
\begin{barticle}[author]
\bauthor{\bsnm{Garlappi},~\bfnm{Lorenzo}\binits{L.}},
  \bauthor{\bsnm{Uppal},~\bfnm{Raman}\binits{R.}} \AND
  \bauthor{\bsnm{Wang},~\bfnm{Tan}\binits{T.}}
(\byear{2007}).
\btitle{Portfolio selection with parameter and model uncertainty: A multi-prior
  approach}.
\bjournal{Review of Financial Studies}
\bvolume{20}
\bpages{41--81}.
\end{barticle}
\endbibitem

\bibitem[\protect\citeauthoryear{Gron, J{\o}rgensen and
  Polson}{2012}]{gron2012optimal}
\begin{barticle}[author]
\bauthor{\bsnm{Gron},~\bfnm{Anne}\binits{A.}},
  \bauthor{\bsnm{J{\o}rgensen},~\bfnm{Bj{\o}rn~N}\binits{B.~N.}} \AND
  \bauthor{\bsnm{Polson},~\bfnm{Nicholas~G}\binits{N.~G.}}
(\byear{2012}).
\btitle{Optimal portfolio choice and stochastic volatility}.
\bjournal{Applied Stochastic Models in Business and Industry}
\bvolume{28}
\bpages{1--15}.
\end{barticle}
\endbibitem

\bibitem[\protect\citeauthoryear{Hahn and Carvalho}{2015}]{HahnCarvalho}
\begin{barticle}[author]
\bauthor{\bsnm{Hahn},~\bfnm{P~Richard}\binits{P.~R.}} \AND
  \bauthor{\bsnm{Carvalho},~\bfnm{Carlos~M}\binits{C.~M.}}
(\byear{2015}).
\btitle{Decoupling shrinkage and selection in Bayesian linear models: a
  posterior summary perspective}.
\bjournal{Journal of the American Statistical Association}
\bvolume{110}
\bpages{435--448}.
\end{barticle}
\endbibitem

\bibitem[\protect\citeauthoryear{Harrison and
  West}{1999}]{harrison1999bayesian}
\begin{bbook}[author]
\bauthor{\bsnm{Harrison},~\bfnm{Jeff}\binits{J.}} \AND
  \bauthor{\bsnm{West},~\bfnm{Mike}\binits{M.}}
(\byear{1999}).
\btitle{Bayesian Forecasting \& Dynamic Models}.
\bpublisher{Springer}.
\end{bbook}
\endbibitem

\bibitem[\protect\citeauthoryear{Jacquier and
  Polson}{2010}]{jacquier2010simulation}
\begin{bmisc}[author]
\bauthor{\bsnm{Jacquier},~\bfnm{Eric}\binits{E.}} \AND
  \bauthor{\bsnm{Polson},~\bfnm{Nicholas}\binits{N.}}
(\byear{2010}).
\btitle{Simulation-based-estimation in portfolio selection}.
\end{bmisc}
\endbibitem

\bibitem[\protect\citeauthoryear{Jacquier and Polson}{2012}]{jacquier2012asset}
\begin{barticle}[author]
\bauthor{\bsnm{Jacquier},~\bfnm{Eric}\binits{E.}} \AND
  \bauthor{\bsnm{Polson},~\bfnm{Nicholas~G}\binits{N.~G.}}
(\byear{2012}).
\btitle{Asset allocation in finance: A bayesian perspective}.
\bjournal{Hierarchinal models and MCMC: a Tribute to Adrian Smith}.
\end{barticle}
\endbibitem

\bibitem[\protect\citeauthoryear{Jagannathan and
  Ma}{2002}]{jagannathan2002risk}
\begin{btechreport}[author]
\bauthor{\bsnm{Jagannathan},~\bfnm{Ravi}\binits{R.}} \AND
  \bauthor{\bsnm{Ma},~\bfnm{Tongshu}\binits{T.}}
(\byear{2002}).
\btitle{Risk reduction in large portfolios: Why imposing the wrong constraints
  helps}
\btype{Technical Report},
\bpublisher{National Bureau of Economic Research}.
\end{btechreport}
\endbibitem

\bibitem[\protect\citeauthoryear{James and Stein}{1961}]{james1961estimation}
\begin{binproceedings}[author]
\bauthor{\bsnm{James},~\bfnm{William}\binits{W.}} \AND
  \bauthor{\bsnm{Stein},~\bfnm{Charles}\binits{C.}}
(\byear{1961}).
\btitle{Estimation with quadratic loss}.
In \bbooktitle{Proceedings of the fourth Berkeley symposium on mathematical
  statistics and probability}
\bvolume{1}
\bpages{361--379}.
\end{binproceedings}
\endbibitem

\bibitem[\protect\citeauthoryear{Jobson and
  Korkie}{1980}]{jobson1980estimation}
\begin{barticle}[author]
\bauthor{\bsnm{Jobson},~\bfnm{J~David}\binits{J.~D.}} \AND
  \bauthor{\bsnm{Korkie},~\bfnm{Bob}\binits{B.}}
(\byear{1980}).
\btitle{Estimation for Markowitz efficient portfolios}.
\bjournal{Journal of the American Statistical Association}
\bvolume{75}
\bpages{544--554}.
\end{barticle}
\endbibitem

\bibitem[\protect\citeauthoryear{Johannes, Korteweg and
  Polson}{2014}]{johannes2014sequential}
\begin{barticle}[author]
\bauthor{\bsnm{Johannes},~\bfnm{Michael}\binits{M.}},
  \bauthor{\bsnm{Korteweg},~\bfnm{Arthur}\binits{A.}} \AND
  \bauthor{\bsnm{Polson},~\bfnm{Nicholas}\binits{N.}}
(\byear{2014}).
\btitle{Sequential learning, predictability, and optimal portfolio returns}.
\bjournal{The Journal of Finance}
\bvolume{69}
\bpages{611--644}.
\end{barticle}
\endbibitem

\bibitem[\protect\citeauthoryear{Jorion}{1985}]{jorion1985international}
\begin{barticle}[author]
\bauthor{\bsnm{Jorion},~\bfnm{Philippe}\binits{P.}}
(\byear{1985}).
\btitle{International portfolio diversification with estimation risk}.
\bjournal{Journal of Business}
\bpages{259--278}.
\end{barticle}
\endbibitem

\bibitem[\protect\citeauthoryear{Jorion}{1986}]{jorion1986bayes}
\begin{barticle}[author]
\bauthor{\bsnm{Jorion},~\bfnm{Philippe}\binits{P.}}
(\byear{1986}).
\btitle{Bayes-Stein estimation for portfolio analysis}.
\bjournal{Journal of Financial and Quantitative Analysis}
\bvolume{21}
\bpages{279--292}.
\end{barticle}
\endbibitem

\bibitem[\protect\citeauthoryear{Kelly~Jr}{1956}]{kelly1956new}
\begin{barticle}[author]
\bauthor{\bsnm{Kelly~Jr},~\bfnm{John~L}\binits{J.~L.}}
(\byear{1956}).
\btitle{A new interpretation of information rate}.
\bjournal{Information Theory, IRE Transactions on}
\bvolume{2}
\bpages{185--189}.
\end{barticle}
\endbibitem

\bibitem[\protect\citeauthoryear{Ledoit and Wolf}{2003a}]{ledoit2003improved}
\begin{barticle}[author]
\bauthor{\bsnm{Ledoit},~\bfnm{Olivier}\binits{O.}} \AND
  \bauthor{\bsnm{Wolf},~\bfnm{Michael}\binits{M.}}
(\byear{2003}a).
\btitle{Improved estimation of the covariance matrix of stock returns with an
  application to portfolio selection}.
\bjournal{Journal of empirical finance}
\bvolume{10}
\bpages{603--621}.
\end{barticle}
\endbibitem

\bibitem[\protect\citeauthoryear{Ledoit and Wolf}{2003b}]{ledoit2003honey}
\begin{barticle}[author]
\bauthor{\bsnm{Ledoit},~\bfnm{Olivier}\binits{O.}} \AND
  \bauthor{\bsnm{Wolf},~\bfnm{Michael}\binits{M.}}
(\byear{2003}b).
\btitle{Honey, I shrunk the sample covariance matrix}.
\bjournal{UPF Economics and Business Working Paper}
\bvolume{691}.
\end{barticle}
\endbibitem

\bibitem[\protect\citeauthoryear{Lintner}{1965}]{Lintner}
\begin{barticle}[author]
\bauthor{\bsnm{Lintner},~\bfnm{John}\binits{J.}}
(\byear{1965}).
\btitle{The valuation of risk assets and the selection of risky investments in
  stock portfolios and capital budgets}.
\bjournal{The review of economics and statistics}
\bpages{13--37}.
\end{barticle}
\endbibitem

\bibitem[\protect\citeauthoryear{Lopes and West}{2004}]{lopes2004bayesian}
\begin{barticle}[author]
\bauthor{\bsnm{Lopes},~\bfnm{Hedibert~Freitas}\binits{H.~F.}} \AND
  \bauthor{\bsnm{West},~\bfnm{Mike}\binits{M.}}
(\byear{2004}).
\btitle{Bayesian model assessment in factor analysis}.
\bjournal{Statistica Sinica}
\bvolume{14}
\bpages{41--68}.
\end{barticle}
\endbibitem

\bibitem[\protect\citeauthoryear{Markowitz}{1952}]{markowitz1952portfolio}
\begin{barticle}[author]
\bauthor{\bsnm{Markowitz},~\bfnm{Harry}\binits{H.}}
(\byear{1952}).
\btitle{Portfolio selection*}.
\bjournal{The journal of finance}
\bvolume{7}
\bpages{77--91}.
\end{barticle}
\endbibitem

\bibitem[\protect\citeauthoryear{Merton}{1980}]{merton1980estimating}
\begin{barticle}[author]
\bauthor{\bsnm{Merton},~\bfnm{Robert~C}\binits{R.~C.}}
(\byear{1980}).
\btitle{On estimating the expected return on the market: An exploratory
  investigation}.
\bjournal{Journal of Financial Economics}
\bvolume{8}
\bpages{323--361}.
\end{barticle}
\endbibitem

\bibitem[\protect\citeauthoryear{Meucci}{2009}]{meucci2009risk}
\begin{bbook}[author]
\bauthor{\bsnm{Meucci},~\bfnm{Attilio}\binits{A.}}
(\byear{2009}).
\btitle{Risk and asset allocation}.
\bpublisher{Springer Science \& Business Media}.
\end{bbook}
\endbibitem

\bibitem[\protect\citeauthoryear{Murray et~al.}{2013}]{murray2013bayesian}
\begin{barticle}[author]
\bauthor{\bsnm{Murray},~\bfnm{Jared~S}\binits{J.~S.}},
  \bauthor{\bsnm{Dunson},~\bfnm{David~B}\binits{D.~B.}},
  \bauthor{\bsnm{Carin},~\bfnm{Lawrence}\binits{L.}} \AND
  \bauthor{\bsnm{Lucas},~\bfnm{Joseph~E}\binits{J.~E.}}
(\byear{2013}).
\btitle{Bayesian Gaussian copula factor models for mixed data}.
\bjournal{Journal of the American Statistical Association}
\bvolume{108}
\bpages{656--665}.
\end{barticle}
\endbibitem

\bibitem[\protect\citeauthoryear{Ng}{1991}]{ng1991tests}
\begin{barticle}[author]
\bauthor{\bsnm{Ng},~\bfnm{Lilian}\binits{L.}}
(\byear{1991}).
\btitle{Tests of the CAPM with time-varying covariances: A multivariate GARCH
  approach}.
\bjournal{The Journal of Finance}
\bvolume{46}
\bpages{1507--1521}.
\end{barticle}
\endbibitem

\bibitem[\protect\citeauthoryear{Pastor and Veronesi}{2009}]{PandV}
\begin{btechreport}[author]
\bauthor{\bsnm{Pastor},~\bfnm{Lubos}\binits{L.}} \AND
  \bauthor{\bsnm{Veronesi},~\bfnm{Pietro}\binits{P.}}
(\byear{2009}).
\btitle{Learning in financial markets}
\btype{Technical Report},
\bpublisher{National Bureau of Economic Research}.
\end{btechreport}
\endbibitem

\bibitem[\protect\citeauthoryear{Pettenuzzo and
  Ravazzolo}{2015}]{pettenuzzo2015optimal}
\begin{barticle}[author]
\bauthor{\bsnm{Pettenuzzo},~\bfnm{Davide}\binits{D.}} \AND
  \bauthor{\bsnm{Ravazzolo},~\bfnm{Francesco}\binits{F.}}
(\byear{2015}).
\btitle{Optimal portfolio choice under decision-based model combinations}.
\bjournal{Journal of Applied Econometrics}.
\end{barticle}
\endbibitem

\bibitem[\protect\citeauthoryear{Polson and Tew}{1999}]{polson1999bayesian}
\begin{barticle}[author]
\bauthor{\bsnm{Polson},~\bfnm{N}\binits{N.}} \AND
  \bauthor{\bsnm{Tew},~\bfnm{B}\binits{B.}}
(\byear{1999}).
\btitle{Bayesian Portfolio Selection: An analysis of the S\&P 500 index
  1970-1996}.
\bjournal{Journal of Business and Economic Statistics}
\bvolume{18}
\bpages{164--173}.
\end{barticle}
\endbibitem

\bibitem[\protect\citeauthoryear{Puelz, Carvalho and
  Hahn}{2015}]{puelz2015optimal}
\begin{barticle}[author]
\bauthor{\bsnm{Puelz},~\bfnm{David}\binits{D.}},
  \bauthor{\bsnm{Carvalho},~\bfnm{Carlos~M}\binits{C.~M.}} \AND
  \bauthor{\bsnm{Hahn},~\bfnm{P~Richard}\binits{P.~R.}}
(\byear{2015}).
\btitle{Optimal ETF Selection for Passive Investing}.
\bjournal{arXiv preprint arXiv:1510.03385}.
\end{barticle}
\endbibitem

\bibitem[\protect\citeauthoryear{Puelz, Hahn and
  Carvalho}{2016}]{puelz2016variable}
\begin{barticle}[author]
\bauthor{\bsnm{Puelz},~\bfnm{David}\binits{D.}},
  \bauthor{\bsnm{Hahn},~\bfnm{P~Richard}\binits{P.~R.}} \AND
  \bauthor{\bsnm{Carvalho},~\bfnm{Carlos}\binits{C.}}
(\byear{2016}).
\btitle{Variable selection in seemingly unrelated regressions with random
  predictors}.
\bjournal{Available at SSRN}.
\end{barticle}
\endbibitem

\bibitem[\protect\citeauthoryear{Sharpe}{1964}]{Sharpe}
\begin{barticle}[author]
\bauthor{\bsnm{Sharpe},~\bfnm{William~F}\binits{W.~F.}}
(\byear{1964}).
\btitle{Capital asset prices: A theory of market equilibrium under conditions
  of risk*}.
\bjournal{The journal of finance}
\bvolume{19}
\bpages{425--442}.
\end{barticle}
\endbibitem

\bibitem[\protect\citeauthoryear{Sharpe}{1966}]{sharpe1966mutual}
\begin{barticle}[author]
\bauthor{\bsnm{Sharpe},~\bfnm{William~F}\binits{W.~F.}}
(\byear{1966}).
\btitle{Mutual fund performance}.
\bjournal{Journal of business}
\bpages{119--138}.
\end{barticle}
\endbibitem

\bibitem[\protect\citeauthoryear{Takeda et~al.}{2013}]{takeda2013simultaneous}
\begin{barticle}[author]
\bauthor{\bsnm{Takeda},~\bfnm{Akiko}\binits{A.}},
  \bauthor{\bsnm{Niranjan},~\bfnm{Mahesan}\binits{M.}},
  \bauthor{\bsnm{Gotoh},~\bfnm{Jun-ya}\binits{J.-y.}} \AND
  \bauthor{\bsnm{Kawahara},~\bfnm{Yoshinobu}\binits{Y.}}
(\byear{2013}).
\btitle{Simultaneous pursuit of out-of-sample performance and sparsity in index
  tracking portfolios}.
\bjournal{Computational Management Science}
\bvolume{10}
\bpages{21--49}.
\end{barticle}
\endbibitem

\bibitem[\protect\citeauthoryear{Tibshirani}{1996}]{Tib}
\begin{barticle}[author]
\bauthor{\bsnm{Tibshirani},~\bfnm{Robert}\binits{R.}}
(\byear{1996}).
\btitle{Regression shrinkage and selection via the lasso}.
\bjournal{Journal of the Royal Statistical Society. Series B (Methodological)}
\bpages{267--288}.
\end{barticle}
\endbibitem

\bibitem[\protect\citeauthoryear{Wang et~al.}{2011}]{wang2011dynamic}
\begin{barticle}[author]
\bauthor{\bsnm{Wang},~\bfnm{Hao}\binits{H.}},
  \bauthor{\bsnm{Reeson},~\bfnm{Craig}\binits{C.}},
  \bauthor{\bsnm{Carvalho},~\bfnm{Carlos~M}\binits{C.~M.}} \betal{et~al.}
(\byear{2011}).
\btitle{Dynamic financial index models: Modeling conditional dependencies via
  graphs}.
\bjournal{Bayesian Analysis}
\bvolume{6}
\bpages{639--664}.
\end{barticle}
\endbibitem

\bibitem[\protect\citeauthoryear{Wu and Yang}{2014}]{Wu2}
\begin{barticle}[author]
\bauthor{\bsnm{Wu},~\bfnm{Lan}\binits{L.}} \AND
  \bauthor{\bsnm{Yang},~\bfnm{Yuehan}\binits{Y.}}
(\byear{2014}).
\btitle{Nonnegative Elastic Net and application in index tracking}.
\bjournal{Applied Mathematics and Computation}
\bvolume{227}
\bpages{541--552}.
\end{barticle}
\endbibitem

\bibitem[\protect\citeauthoryear{Wu, Yang and Liu}{2014}]{Wu}
\begin{barticle}[author]
\bauthor{\bsnm{Wu},~\bfnm{Lan}\binits{L.}},
  \bauthor{\bsnm{Yang},~\bfnm{Yuehan}\binits{Y.}} \AND
  \bauthor{\bsnm{Liu},~\bfnm{Hanzhong}\binits{H.}}
(\byear{2014}).
\btitle{Nonnegative-lasso and application in index tracking}.
\bjournal{Computational Statistics \& Data Analysis}
\bvolume{70}
\bpages{116--126}.
\end{barticle}
\endbibitem

\bibitem[\protect\citeauthoryear{Yen}{2013}]{yen2013note}
\begin{btechreport}[author]
\bauthor{\bsnm{Yen},~\bfnm{Yu-Min}\binits{Y.-M.}}
(\byear{2013}).
\btitle{A note on sparse minimum variance portfolios and coordinate-wise
  descent algorithms}
\btype{Technical Report}.
\end{btechreport}
\endbibitem

\bibitem[\protect\citeauthoryear{Yen and Yen}{2014}]{yen2014solving}
\begin{barticle}[author]
\bauthor{\bsnm{Yen},~\bfnm{Yu-Min}\binits{Y.-M.}} \AND
  \bauthor{\bsnm{Yen},~\bfnm{Tso-Jung}\binits{T.-J.}}
(\byear{2014}).
\btitle{Solving norm constrained portfolio optimization via coordinate-wise
  descent algorithms}.
\bjournal{Computational Statistics \& Data Analysis}
\bvolume{76}
\bpages{737--759}.
\end{barticle}
\endbibitem

\end{thebibliography}

\end{document}